\documentclass[final,5p,numbers,times,twocolumn]{elsarticle}

\usepackage[english]{babel}
\usepackage{bibentry}
\usepackage[utf8]{inputenc}
\usepackage{listings}

\usepackage{algorithm}
\usepackage{algpseudocode}

\algrenewcommand{\alglinenumber}[1]{\color{gray!80!gray}\tiny#1:}

\usepackage[table,usenames,dvipsnames]{xcolor}
\usepackage{colortbl}
\usepackage{color}
\definecolor{lightgray}{rgb}{.9,.9,.9}
\definecolor{darkgray}{rgb}{.4,.4,.4}
\definecolor{light-gray1}{gray}{0.92}
\definecolor{light-gray2}{gray}{0.84}

\usepackage[colorlinks,bookmarksopen,bookmarksnumbered,citecolor=red,urlcolor=red]{hyperref}

\usepackage{amssymb}

\newcommand\BibTeX{{\rmfamily B\kern-.05em \textsc{i\kern-.025em b}\kern-.08em
T\kern-.1667em\lower.7ex\hbox{E}\kern-.125emX}}

\journal{Journal of Systems and Software}

\begin{document}

\begin{frontmatter}

\title{Developing a Model-Driven Reengineering Approach for Migrating
    PL/SQL Triggers to Java: A Practical Experience}

\author[mur]{Carlos J. Fernandez-Candel}
\ead{cjferna@um.es}
\author[mur]{Jesus Garcia-Molina}
\ead{jmolina@um.es}
\author[mur]{Francisco Javier Bermudez Ruiz}
\ead{fjavier@um.es}
\author[mur]{Jose Ramon Hoyos Barcelo}
\ead{jose.hoyos@um.es}
\author[mur]{Diego Sevilla Ruiz}
\ead{dsevilla@um.es}
\author[ope]{Benito Jose Cuesta Viera}
\ead{bcuesta@opencanarias.es}
\address[mur]{Faculty of Informatics, University of Murcia, Spain}
\address[ope]{Open Canarias S.L., Spain}

\begin{abstract}
  Model-driven software engineering (MDE) techniques are not only useful in
  forward engineering scenarios, but can also be successfully applied to
  evolve existing systems. RAD (Rapid Application Development) platforms
  emerged in the nineties, but the success of modern software technologies
  motivated that a large number of enterprises tackled the migration of
  their RAD applications, such as Oracle Forms.
  Our research group has collaborated with a software company in developing
  a solution to migrate PL/SQL monolithic code on Forms triggers and
  program units to Java code separated in several tiers.

  Our research focused on the model-driven reengineering process applied to
  develop the migration tool for the conversion of PL/SQL code to Java.
  Legacy code is represented in form of KDM (Knowledge-Discovery Metamodel)
  models. In this paper, we propose a software process to implement a
  model-driven re-engineering. This process integrates a TDD-like approach
  to incrementally develop model transformations with three kinds of
  validations for the generated code. The implementation and validation of
  the re-engineering approach are explained in detail, as well as the
  evaluation of some issues related with the application of MDE.
\end{abstract}


\end{frontmatter}

\section{Introduction\label{sec:Introduction}}

Model-driven software engineering (MDE) techniques are not only useful in
forward engineering scenarios, but can also be successfully applied to
evolve existing systems. Models are very appropriate to represent
information involved in evolution tasks at a higher level of abstraction
(e.g. information harvested into a reverse engineering process). Moreover,
model transformation technologies have been developed to facilitate the
automation of such tasks.

RAD (Rapid Application Development) platforms emerged in the nineties to
offer an agile way of developing software. Agility was the result of
applying a UI-centered paradigm in which most application code was
entangled in the event handlers. Therefore, the gain of productivity was
achieved at the expense of software quality, and maintenance was negatively
affected~\cite{andrade2008}. The success of the object-oriented paradigm and the appearance of
modern software technologies motivated that a large number of enterprises
tackled the migration of their RAD applications. This was mainly due to the
fact that most of the RAD environments were discontinued.
Oracle Forms has been one of the most widely used RAD technologies over
last three decades. Although the Oracle company continues supporting and
offering solutions to integrate Forms with new technologies (e.g. Web and
Java), many enterprises around the world have migrated its Forms legacy
code to modern platforms in order to take advantage of new software
technologies.

Open Canarias is a Spanish software company with years of experience in
software modernization, specially in the banking area. In~2017, Open
Canarias launched the Morpheus project\footnote{This project has been
  partially supported with funds of the Spanish Industry Ministry through
  the CDTI project: \textbf{IDI-20150952} (\url{http://www.cdti.es/}).} to
develop a tool able of achieving the highest degree of automation possible
in the migration of Oracle Forms applications to object-oriented platforms.
An MVC architecture based on Java frameworks was the selected target
platform. Our research group collaborated with Open Canarias in developing
a solution to migrate PL/SQL code on triggers and program units to Java
code. This paper is focused on the model-driven approach devised to tackle
such a migration.

A software migration is a form of modernization in which an existing
application is moved to a new platform that offers more benefits (e.g.
better maintainability or new functionality.) A re-engineering strategy is
commonly applied to migrate systems in a systematic
way~\cite{Tilley95,kazman98}. This strategy consists of three stages.
Firstly, a reverse engineering state is carried out to obtain a
representation of the legacy system at a higher level of abstraction. This
representation is mapped to the new architecture in a second stage.
Finally, a forward engineering stage generates the target artefacts of the
new application from the representation generated in the restructuring
stage~\cite{kazman98}. Models and model transformations can be used to
implement the transformation processes involved in these three stages as
illustrated in~\cite{javi2017-is,ase10}. Actually, model-driven
re-engineering and model-driven reverse engineering are common application
scenarios of MDE techniques~\cite{jordibook}. Some of the most significant
efforts made in model-driven software modernization can be found in a
recently published systematic literature review on model-driven reverse
engineering approaches~\cite{Raibulet2017}.

In Morpheus, we have devised a model-driven re-engineering process for the
migration from PL/SQL to Java. KDM (Knowledge-Discovery
Metamodel)~\cite{kdm} is used to represent the PL/SQL legacy code.
PL/SQL triggers mix concerns that must be separated in three tiers in the
target MVC architecture. This separation of concerns was one of the main
challenges to be addressed in our reengineering process. To facilitate the
code separation and achieve a language-independent solution, we have used
the notion of \emph{primitive operation} proposed in a previous work of our
group~\cite{wcre11}. KDM models are transformed in models that represent
PL-SQL trigger code in form of primitive operations commonly used in
writing event handlers of RAD applications. The primitive operations models
obtained have been used to generate a model that represents how legacy code
is separated into the three tiers of an MVC architecture. In turn, these
target platform models are transformed into Object-Oriented models before
generating Java code of the new application.

We have defined a development process that involves two main stages. First,
individual model transformations are incrementally implemented following a
strategy inspired on test-first approach.
Once the model transformation chain has been
completed, several kinds of tests are applied on the generated code.
The migration tool has been validated by means of forms provided by Open
Canarias, which are part of real applications.
In this paper, we will describe in detail each stage of both the
re-engineering solution proposed and the validation process carried out.
For each model transformation, we will discuss how it has been implemented
and tested. A trigger example will be used to illustrate how each
transformation works. Moreover, we evaluate some key aspects of a MDE
solution in the migration scenario here considered
such as~(i)~the adoption of KDM;~(ii)~the choice of the language for
writing model-to-model transformations;~(iii)~the difficulties encountered
for testing transformations. Moreover, we will propose a model
visualization technique based on the conversion of models into graph
database instances.

\paragraph{Contributions}

Model-driven reengineering approaches for migrating monolithic legacy code
to~3-tier architectures have been presented in~\cite{wcre11}
and~\cite{andrade2008}. These works are mainly focused on the reverse
engineering stage. Moreover, they do not pay attention to the development
process applied to implement the re-engineering. In particular, how each
model transformation and the final solution were validated is not
addressed.

Writing model transformations is recognized as a challenging task in
developing model transformations~\cite{baudry2010}, and some approaches to
develop transformation chains~\cite{kuster2009} or individual
transformations~\cite{Varro} have been proposed. However, experiences on
real projects in developing model transformations 
are not published as far as we know. Test-driven development
is also an idea not yet sufficiently explored for individual
transformations, although it has been proposed for incrementally developing
transformation chains~\cite{kuster2009}.

A limited number of industrial experiences of model-driven modernization
have been reported so far. This is evidenced by the low number of
model-driven reverse engineering works obtained in the systematic
literature review presented in~\cite{Raibulet2017}. Moreover,
only three of the eleven papers considered in that review used the standard
KDM metamodel. To our knowledge, no work on a source-to-source conversion
based on KDM has been published so far. This may lead to think that there
is a lack of interest in applying MDE techniques in software modernization.
A dissemination of knowledge on real experiences may favor a growth in the
usage of model-driven modernization if benefits are reported.

Several commercial and open source tools are available to migrate PL/SQL
triggers to the Java platform, such as \emph{Ispirer
  MnMTK}\footnote{\url{http://www.ispirer.com}.} and
\emph{PLSQL2Java}.\footnote{\url{https://pitss.com}.} These tools convert
PL/SQL code into Java, but the separation of tiers is not considered.

With these premises, we introduce the main research contributions of this
work:

\begin{itemize}

\item A development process for implementing a re-engineering is defined
  and applied. All the stages of a re-engineering are addressed.
\item A test-driven strategy to incrementally develop model transformations
  in a software migration is proposed.
\item A comparison between Java and two widely used transformation
  languages (ATL~\cite{atl} and QVT operational~\cite{qvt}) is presented.
\item A novel strategy for visualizing models using graph databases to help
  with testing is shown.
\item KDM is used and evaluated in a scenario of code migration.

\end{itemize}

It should also be noted that a valuable contribution of our work is the
ability to develop a model-driven migration solution to be integrated in a
tool developed by a company, which covers the design, implementation, and
validation stages as well and evaluation of the practical experience.

\paragraph{Paper organization}

Section~\ref{sec:background} introduces the background needed to better
understand the paper: the notion of model-driven reengineering and the KDM
metamodel. Section~\ref{sec:Requirements} states the requirements and
challenges in building the migration tool. Section~\ref{sec:Example}
presents the migration problem through a trigger example and shows how the
trigger code is separated in several tiers in Java.
Section~\ref{sec:Overview} shows a overview 
of the model-driven reengineering process
applied to implement the tool.
Section~\ref{sec:Development} explains the development process defined to
implement the solution. Section~\ref{sec:Implementation} explains how each
stage of the re-engineering process has been implemented and tested, and
describes all the metamodels involved. In addition, the trigger example
introduced in Section~\ref{sec:Example} is used to show how each
transformation works. Section~\ref{sec:Validation} explains how the tool
has been validated. Section~\ref{sec:Evaluation} evaluates some MDE-related
issues from the experience gained in our work.
Section~\ref{sec:RelatedWork} discusses how our work is related to other
proposals. Section~\ref{sec:Conclusions} draws some conclusions and
comments on future works.

\section{Background\label{sec:background}}

This section is devoted to introduce some foundations for a better
understanding of this paper. First, we will present the notion of
model-driven re-engineering, and afterwards we will explain the elements of
the KDM metamodel involved in the proposed migration solution.

\subsection{Model-driven reengineering process\label{sec:horseshoe}}

\begin{figure}[!h]
  \centering
  \includegraphics[width=0.95\columnwidth]{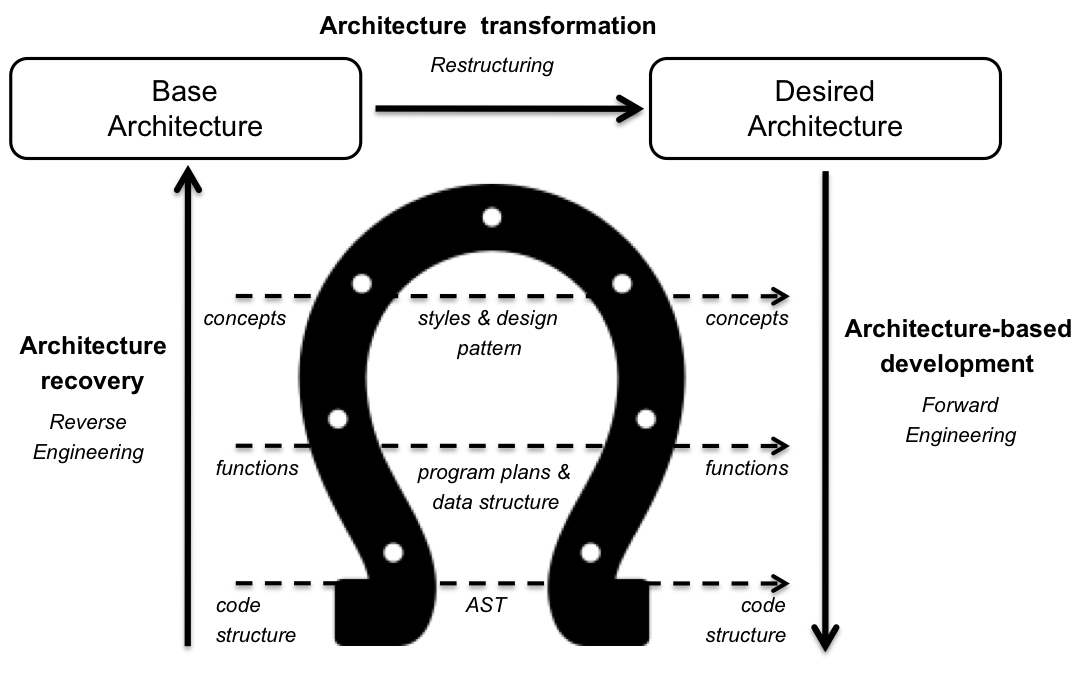}
  \caption{Horseshoe model for re-engineering (\cite{kazman98}).}
  \label{fig:horseshoe}
\end{figure}

Software re-engineering offers a disciplined way to migrate a legacy
system~\cite{Tilley95}. Three kind of activities are normally performed in
a re-engineering: reverse engineering, forward engineering, and
restructuring~\cite{chikofsky1990}. Figure~\ref{fig:horseshoe} shows the
horseshoe model~\cite{kazman98} that is frequently used to illustrate how
reengineering approaches work. Reverse engineering is applied on the
existing system in order to harvest knowledge in form of abstract
representations. Through a chain of vertical transformations, descriptions
of several system aspects are extracted and represented at different levels
of abstraction. The new source code is created by means of vertical
transformations (forward engineering) whose input are representations
defined for aspects of the target system. These representations are
obtained through horizontal transformations (restructuring) whose input is
a representation at the same level of abstraction produced in the reverse
engineering process. In a re-engineering process, transformations can be
either manually carried out by developers or automatically executed by
means of specific programs.

\begin{figure}[!h]
  \centering
  \includegraphics[width=0.95\columnwidth]{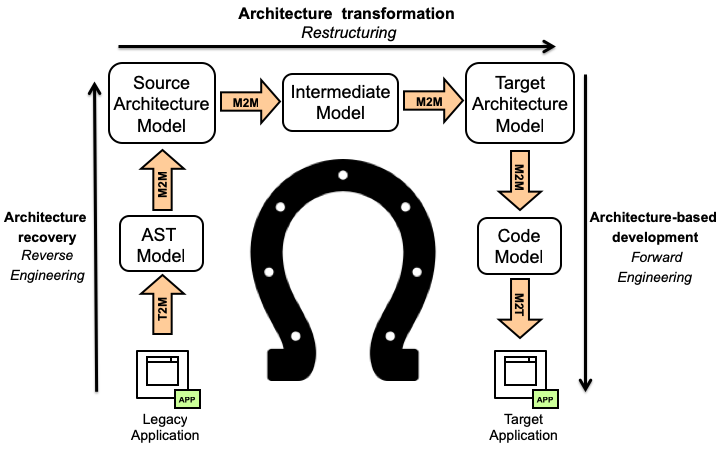}
  \caption{A scenario of model-driven reengineering.}
  \label{fig:mde-horseshoe}
\end{figure}

Creating abstract representations and writing transformations between
representations are, therefore, essential activities in re-engineering
processes. This has motivated the emergence of model-driven re-engineering
approaches, which take advantage of MDE techniques~\cite{jordibook},
specially (meta)models and model transformations, in order to automate such
processes. Metamodels are used to formally define the different
representations, and vertical and horizontal transformations are automated
with model transformations: \emph{text-to-model transformations} (t2m) to
extract models from code, \emph{model-to-text transformations} (m2t) to
generate code from models, and \emph{model-to-model transformations} (m2m)
when the input and output are models. A possible scenario of model-driven
engineering is shown in Figure~\ref{fig:mde-horseshoe}. Throughout the
paper, the term ``injection'' will be used to refer to the process of
extracting models from source code, as it is common in MDE
literature~\cite{jordibook}.

\subsection{KDM metamodel\label{sec:KDM}}

\begin{figure*}[!t]
\centering\includegraphics[width=0.90\linewidth]{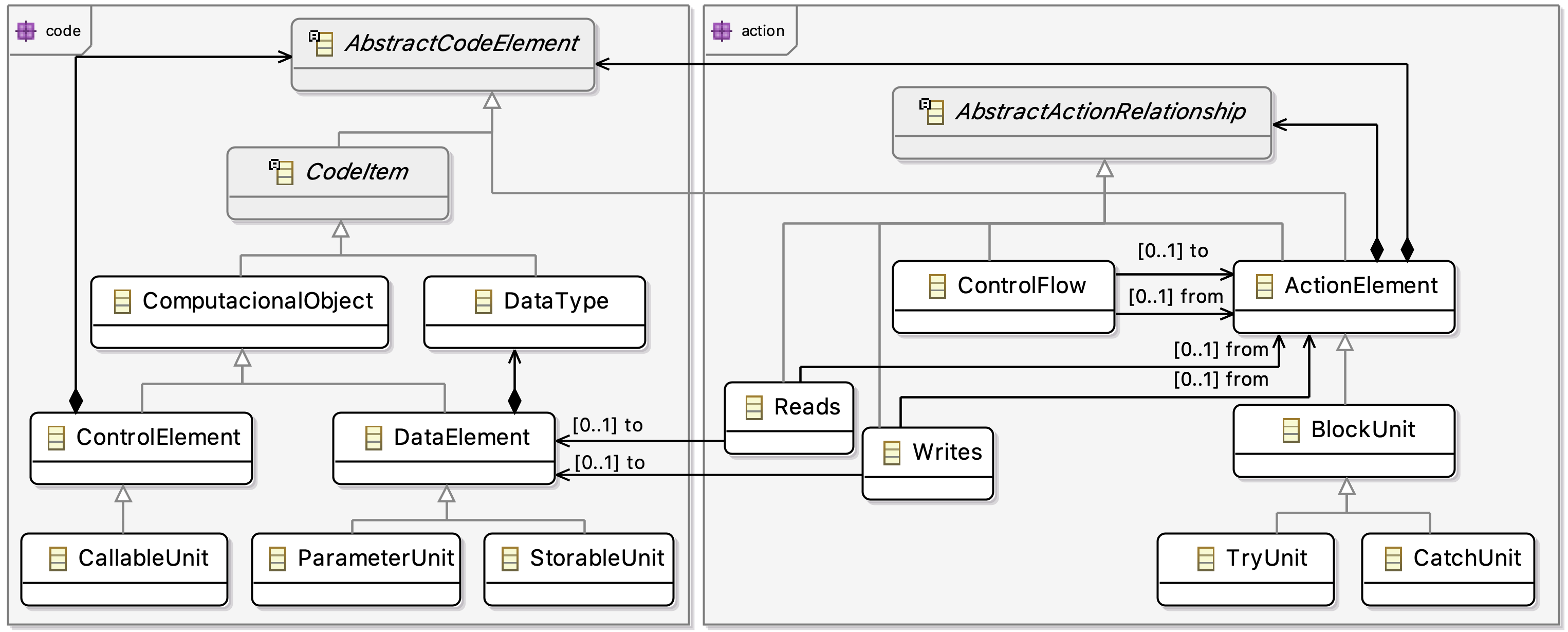}
\caption{An excerpt of the \texttt{Code} and \texttt{Action} packages.}
\label{fig:Code-Action}
\end{figure*}

Architecture-Driven Modernization (ADM) is an OMG initiative to support
modernization using model-driven techniques. ADM was launched in~2003 with
the main purpose of providing a set of metamodels aimed to represent the
information commonly used in software modernization tasks.
KDM is the core metamodel of ADM. It is a language-independent metamodel
that permits the representation of legacy software assets at different
levels of abstraction, ranging from source code to higher-level
abstractions such as UI events, platforms, or business rules. KDM models
can be built from abstract syntax tree (AST) models which conform to the
Abstract Syntax Tree Metamodel (ASTM)~\cite{astm}. The rest of metamodels
and specifications of ADM are based on KDM, such as SMM~\cite{smm} for
representing metrics, and AFP~\cite{AFP} for automating the function points
counting. Next, we will briefly describe how KDM is organized in packages
and layers, and then the packages used in our solution to represent PL/SQL
code will be explained in more detail: \texttt{Action}, \texttt{Code}, and
\texttt{UI}.

\subsubsection{KDM overview\label{kdm-overview}}

KDM is a large metamodel organized in twelve packages which relate to each
other. These packages are partitioned in four layers that represent
different domains of a software system. The \emph{Infrastructure} layer
includes the \texttt{Core}, \texttt{KDM}, and \texttt{Source} packages,
which provide the basic concepts that are used in the rest of levels. The
\emph{Program Elements} layer defines constructs for commonly used
programming languages: organizational and descriptive elements (e.g.,
types, modules, classes, and procedures) in the \texttt{Code} package, and
behavioral elements (e.g., statements and control flow) in the
\texttt{Action} package. The \emph{Resource} layer includes packages to
describe resources managed at runtime, in particular data, user interfaces,
events, and platforms. Finally, the \emph{Abstraction} layer deals with the
architectural view, the domain conceptual modeling, and the build process
of the system. Notice that the information related to \emph{Infrastructure}
and \emph{Program Elements} can be directly extracted from the source code,
but models for the other packages must be inferred from such information.

The KDM specification also offers an \emph{extension mechanism}, where an
extension is defined as a \emph{family of stereotypes}. Each stereotype has
a name and aggregates a set of {\em tags} (pairs name-type) that express
the properties or attributes characterizing it. KDM is a
language-independent specification, but modeling source code requires to
represent the precise semantic of each statement. For this, KDM provides a
set of micro-actions, named \emph{Micro-KDM}, which can be thought of as
equivalent to an intermediate representation. These micro-actions provide
precise semantics for the basic operations in general programming
languages, or GPLs, such as comparison, control, and operations on
primitive types. Instead of using Micro-KDM, a creator of KDM models for a
particular GPL could define a stereotype family, but then these models
could not be interchanged with other KDM-based tools.

KDM emerged as a common interchange format to favor interoperability and
data exchange among modernization tools developed by different vendors.
Three levels of conformance are defined in the KDM specification. Each
level establishes the packages that a tool should support to be compliant
with that level. A tool is \emph{L0~compliant} if it supports packages in
\emph{Infrastructure} and \emph{Program Elements} layers. If a tool is
L0~compliant and also support a package included in the \emph{Resources}
and \emph{Abstractions} layers, then it is \emph{L1~compliant for the
  corresponding package} (e.g. \texttt{Data} or \texttt{UI}). Finally, L2
level requires to be compliant with all the packages that are part of the
\emph{Resources} and \emph{Abstractions} layers.

\subsubsection{Code and Action packages\label{code-action-kdm}}

An excerpt of the \texttt{Code} package appears on the left of
Figure~\ref{fig:Code-Action}. \texttt{CodeItem} is the central element of
this package. It represents any element of the source code.
\texttt{ComputationalObject} and \texttt{DataType} inherit from
\texttt{CodeItem}.
\texttt{ComputationalObject} is the root of the \texttt{DataElement} and
\texttt{ControlElement} hierarchies that represent data and control
elements, respectively. \texttt{ControlElement}s represent callable
elements (\texttt{CallableUnit}) such as procedures and methods, and
\texttt{DataElement} represents data items such as variables
(\texttt{StorableUnit}), and parameters (\texttt{ParameterUnit}).
\texttt{DataType} is the root of the class hierarchy that represents data
types (primitive, enumerated, composite, and derived.)

The \texttt{Action} package represents behavior through statements,
conditions, code flows, exceptions, and data readings and writings. An
excerpt of this package appears on the right of
Figure~\ref{fig:Code-Action}. As shown in that figure, the \texttt{Action}
package is defined by extending and referring to elements in the
\texttt{Code} package.
\texttt{ActionElement} is the central element of this package. Both
\texttt{ActionElement} and \texttt{CodeItem} inherit from the abstract
class \texttt{AbstractCodeElement}, included in the Code package.
\texttt{ActionElement} is provided to describe the basic unit of behaviour.
An \texttt{ActionElement} aggregates instances of classes that define
relationships whose source is an \texttt{ActionElement}.
\texttt{AbstractActionRelationship} is the root of the class hierarchy that
represents relationships such as \texttt{ControlFlow}, \texttt{Writes}, and
\texttt{Reads}. An \texttt{ActionElement} object can aggregate instances of
these relationship classes, and each relationship instance will have a
reference to another \texttt{ActionElement} object that describes the code
block where the execution flow will continue. \texttt{BlockUnit} is a
subclass of \texttt{ActionElement} that represents a block of
\texttt{ActionElement}s, for example a block of statements in a container
as a method or trigger. \texttt{ExceptionUnit} is the root of classes for
elements of an exception handler (\texttt{TryUnit}, \texttt{CatchUnit}, and
\texttt{FinallyUnit}.) The \texttt{ExceptionUnit} class, which inherits
from \texttt{BlockUnit}, is not shown for the sake of clarity.

\subsubsection{UI package\label{ui-kdm}}

\begin{figure}[h!]
\centering\includegraphics[width=\columnwidth]{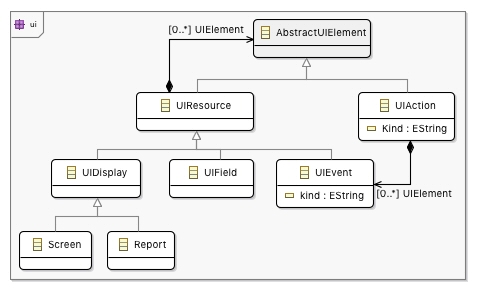}
\caption{An excerpt of the UI Package.}
\label{fig:UIpackage}
\end{figure}

The \texttt{UI} package contains elements that represent several aspects of
the user interface such as the view structure, the data flow, and the
events triggered from visual components. This package depends on the
\texttt{Code} and \texttt{Action} packages commented above.
Figure~\ref{fig:UIpackage} shows the UI elements that represent resources
of a user interface: screen, report, field, event, and action.

\section{Code Migrator tool: Requirements and
  Challenges\label{sec:Requirements}}

In Morpheus, Open Canarias organized the migration tool into two main
components that correspond to the two main artifacts of an Oracle Forms
application: UIs and event handlers. We refer to these components as
\emph{UI Migrator tool} and \emph{Code Migrator tool}, respectively. Our
research group has collaborated in the implementation of the Code Migrator
tool.
In this section, we shall precisely identify the requirements to be
satisfied and the challenges to be addressed in building the tool. In the
following sections, the model re-engineering strategy designed to carry out
the migration and the process defined to implement this strategy will be
presented.

\begin{figure}[!ht]
  \centering
  \includegraphics[width=\columnwidth]{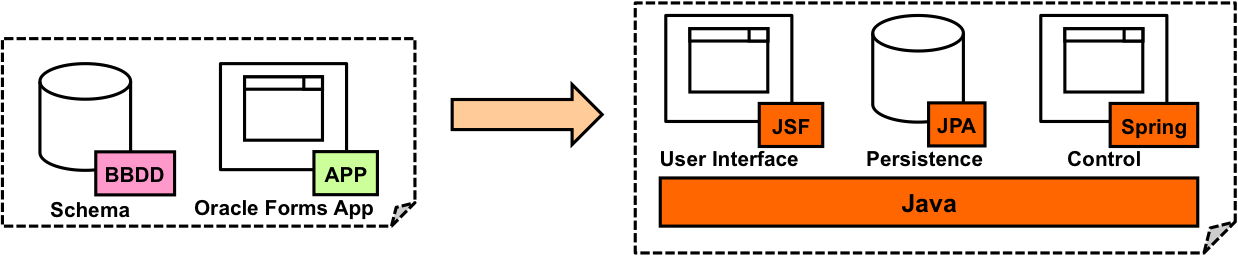}
  \caption{A tool for migrating Oracle Forms applications to Java.}
  \label{fig:esquema_problema}
\end{figure}

As depicted in Figure~\ref{fig:esquema_problema}, the input of the tool is
the source code and the database schema of the Forms application to be
migrated. The output is made up of artifacts generated for a MVC
architecture. In the current implementation, the MVC architecture is based
on Java frameworks: JSF~\cite{jsf} on the view layer,
Spring\footnote{\url{https://spring.io/}.} on the business logic layer, and
JPA~\cite{jpa} on the persistence layer.

The main requirements for the tool are:

\begin{itemize}
\item It should translate procedural PL/SQL code into Java code. Because
  Forms PL/SQL triggers are monolithic, the code generated must be divided
  into several methods according to the tiers of the target platform.

\item \emph{Platform-independence} should be achieved in the design of the
  tool. It should be designed to facilitate generating code for a wide
  range of object-oriented target platforms, e.g., C\# for .NET.

\item A level of automation close to~100\% should be achieved on migrating
  PL/SQL triggers and program units to Java. The level of automation will
  be measured by the percentage of legacy PL/SQL code translated to Java.

\item Those PL/SQL built-in functions/procedures that 
do not have an equivalent method in the Java API
will be manually migrated as part of a library available for all the
  migration projects of the company.

\item The code generated should be integrated into the component that
  automates the UI migration.

\item Not only the trigger code should not be migrated, but also the code
  dealing with database access, and the code of procedures and functions
  invoked from triggers as program units.
\end{itemize}

Next, we will identify the main issues to be addressed to satisfy the
requirements above exposed.

\paragraph{PL/SQL-to-Java mapping} A mapping between the PL/SQL and Java
languages must be established. To achieve a level of~100\% automation on
the PL/SQL code injected
this mapping should cover all the PL/SQL elements: statements, expressions,
data types, cursors, exceptions, and collections. Built-in functions and
predefined exceptions are two examples of PL/SQL elements for which a
direct translation is not possible. Built-in operations should be
implemented directly in Java. Regarding exceptions, Oracle Forms provides a
high number of platform-specific exceptions that must be migrated in two
ways: defining an equivalent Java exception, or either omitting those not
applicable.

\paragraph{Separation of concerns}
As indicated above, event handlers of RAD applications mix code of
different aspects of an application (data access, control logic, and
business logic.) Conversely, \emph{separation of concerns} is a key
architectural design directive in modern paradigms. Disentangling the code
of PL/SQL triggers is undoubtedly the main challenge in implementing our
tool. The code must be analyzed to apply a separation according the tiers
or layers of the target platform, in our case, an MVC architecture. The
code is separated in fragments which must be categorized as belonging to a
particular tier of the target architecture. Once categorized, the code
fragments can be translated and included in a method of a tier.

\paragraph{Sharing variables among generated methods}
A local variable of a PL/SQL function or procedure must be shared among the
Java methods generated for such a function or procedure. A possible
solution would be passing values as arguments in method invocations, but
arguments are always passed by value in Java, so a different solution must
be devised.

\paragraph{Target platform independence}

Metamodels should be defined to facilitate the adaptation of the solution
to different source and target platforms. For example, legacy code can be
expressed in a form that is independent of a particular RAD platform, a
object-oriented metamodel could be defined to ease the code generation for
different object-oriented languages, and an MVC architecture metamodel
would provide independence of a particular target platform.

\paragraph{Integrating the code generated for the UI and the event handlers}
In our case, we had to address the issue of integrating the code generated
for PL/SQL triggers and unit programs with the code previously generated by
the UI Migrator tool.

\section{A migration example\label{sec:Example}}

In this section, we introduce the migration problem to be addressed by
using an example code to be migrated. First, we briefly describe the
structure of a Forms application. Next, we show a trigger and program unit
example. Then, we present the target architecture and explain how the
PL/SQL monolithic code is separated in several tiers. Finally, we show the
Java code that should be generated by the migration process.

\paragraph{Forms applications}
An Oracle Forms application is composed of one or more forms. Each form
contains the artifacts that implement both the user interface and the
business and application logic. A form includes a set of visual components
(i.e. widgets) that belong to one or more windows. Most of the application
code is implemented as triggers that handle visual component events. These
triggers are written in PL/SQL, and can be viewed as the typical event
handler in similar RAD architectures. Apart from triggers, another Form
construct that uses PL/SQL code is a Program Unit. A Program Unit is an
auxiliary function or procedure that provides support for implementing
business and application logic. They can be invoked from triggers.

\paragraph{A PL/SQL trigger example \label{code-example}}

\begin{figure}[!h]
  \centering
  \includegraphics[width=\columnwidth]{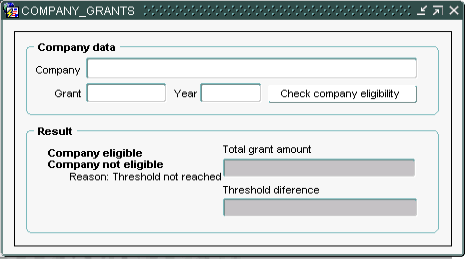}
  \caption{Grants window example.}
  \label{fig:ventana-ejemplo}
\end{figure}

Figure~\ref{fig:ventana-ejemplo} shows a window that is part of an simple
form that was created to provide a trigger and program unit example. This
form could be part of an application used to manage grants for companies.
The window has a button ``{\em Check company eligibility}''. Our form
includes a trigger to manage the {\em pressed} event of such button.
This trigger includes database access and
operations that read or write information on the visual components of a
window. It supports the functionality of applying a new grant
when funds of a previous grant have been spent.
A new grant can only be applied to if the payments made under the previous
grant exceed the threshold indicated in the request. Therefore, the trigger
has to read the payments and the threshold value from the database. Then
all the payments are added to calculate the paid money total. If this total
amount is greater than the threshold, then~(i)~the old grant is
updated,~(ii)~a new grant is created, and~(iii)~the result of the operation
is shown in an application window.

Following, an excerpt of the trigger code is shown:

\begin{lstlisting}[language=SQL,numbers=left,numberstyle=\tiny,morekeywords={BEGIN,END,IF,RAISE}]
BEGIN
  BEGIN
    company_name := normalize_company_name(:COMPANY);
    SELECT sum(PAYMENT) INTO money_paid FROM GRANTS.GRANTS_PAYMENTS WHERE ...;
    SELECT threshold, endowment INTO threshold, endowment FROM GRANTS.COMPANY_GRANTS WHERE ...;

    total := 2 * endowment - money_paid;

    IF money_paid >= threshold THEN
      UPDATE GRANTS.COMPANY_GRANTS_GRANTED SET state = 'SUSPENDED' WHERE ...;
      INSERT INTO GRANTS.COMPANY_GRANTS_GRANTED (...) VALUES (:GRANT_CODE, company_name...);
    END IF;
  EXCEPTION WHEN OTHERS THEN
    message('Database unaccesible');
    RAISE FORM_TRIGGER_FAILURE;
  END;

  IF money_paid >= threshold THEN
    SET_ITEM_PROPERTY('...GRANT_RENEWED', visible, property_true);
  ELSE
    SET_ITEM_PROPERTY('...THRESHOLD_NOT_EXCEEDED', visible, property_true);
  END IF;

  diference := threshold - money_paid;
  IF diference > 0 THEN
    :RENEW_COMPANY_GRANTS.THRESHOLD_DIFERENCE := diference;
  ELSE
    :RENEW_COMPANY_GRANTS.TOTAL_AMOUNT := 2 * endowment - money_paid;
  END IF;
END;
\end{lstlisting}

Note that the trigger code mixes application logic (lines 9, 18, and 24),
database access (lines 4, 5, 10, and 11) and UI reading/writing operations
(lines 19, 21, 26, and 28). The trigger calls the function
\texttt{normalize\_company\_name()} that is part of a program unit.

\begin{lstlisting}[language=SQL,numbers=left,numberstyle=\tiny,morekeywords={BEGIN,END,IF,FUNCTION,RETURN}]
FUNCTION normalize_company_name (company_name IN VARCHAR2) RETURN VARCHAR2 IS
BEGIN
  IF length(company_name) > 256 THEN 
  	RETURN substr(company_name, 1, 256); 
  END IF;
  RETURN company_name;
END;
\end{lstlisting}

In addition, two PL/SQL built-in functions are used: \texttt{length()} that
returns the size of a \texttt{Varchar} variable, and \texttt{substr()} that
returns a substring of a \texttt{Varchar} variable.

\paragraph{MVC target architecture}
Next, we describe the target architecture and explain how the trigger code
is separated in three tiers. This information will help to understand the
implementation of our approach explained in in
Section~\ref{sec:Implementation}.

According to the requirements of Open Canarias, the Java application to be
generated from the legacy code should be organized in accordance with a MVC
architecture based on Java technologies. In particular, the company
requires three Java frameworks to be stacked to form the MVC architecture:
JSF ({\em Java ServerFaces\/}) for the view layer, Spring for organizing the
code of the application logic and supporting the Controller layer, and JPA
({\em Java Persistence API\/})~\cite{jpa} for providing persistence/access
to business entities.

In Spring MVC, each controller is annotated by the \texttt{@Controller}
tag. The navigation among views is given by the use of controllers and
their methods annotated as \texttt{@RequestMapping}. The application logic
is implemented in services classes (also annotated as \texttt{@Service}),
and they are declared in controllers by the injection dependency technique
through the \texttt{@Autowired} declaration.

JSF allows declaring the view. It enables programmers to implement the
view in two separated artefacts:~(i)~a \emph{view} that is composed of UI
widgets, which is implemented as a JSP or Facelets file,
and~(ii)~a \emph{Managed Bean} that provides view backing for the UI
widgets, which is implemented as a Java class. Accessing widget values is
achieved through attribute declarations and their corresponding
getter/setter methods. Widget events are handled through Java methods of
the Managed Bean class. These event handler methods usually contains UI
logic that manipulates the widgets in the view, and they delegate the
application logic and data access on the application controllers. 
This separation was required by Open Canarias. Below, we explain the 
separation strategy applied.

\paragraph{Code separation strategy\label{Code-separation}}

PL/SQL code is separated in the following three kinds of Java classes:

\begin{itemize}
\item A JSF Managed Bean class for each window in the input form. This
  class contains a method for each trigger associated to visual components
  in the window. These trigger methods manipulate UI elements in order to
  read the values introduced by users or showing values which are
  calculated or read from database.
\item A Spring service class that implements an application controller.
  There will be one application controller for each window in the input
  form. These controller classes contain the methods in which the code of a
  legacy trigger is divided as explained below. These methods implement the
  business logic, which contains data access and application logic. In
  addition, a controller class contains methods for some database
  operations, such as insert and update.
\item A Spring service class is created to hold the methods resulting of
  the Program Unit migration. There will be one Spring service class for
  each form which will be in charge of providing all the Java methods
  resulting of the program unit migration.
\end{itemize}

The code of a trigger is separated into methods as follows. Statements that
include write or update operations on elements of the user interface are
mapped to Java and added to the trigger method in the managed bean class.
Code that does not modify UI elements is migrated to a set of methods in
the application controller. There will be one method for each code fragment
that is located before or after a sequence of UI modifying statements.
Whenever a method is created in the controller, a call to such a method is
added to the managed bean just after the last UI modifying code. In
Section~\ref{object}, this code separation is explained in more detail.

\begin{figure}[!t]
 \includegraphics[width=\linewidth]{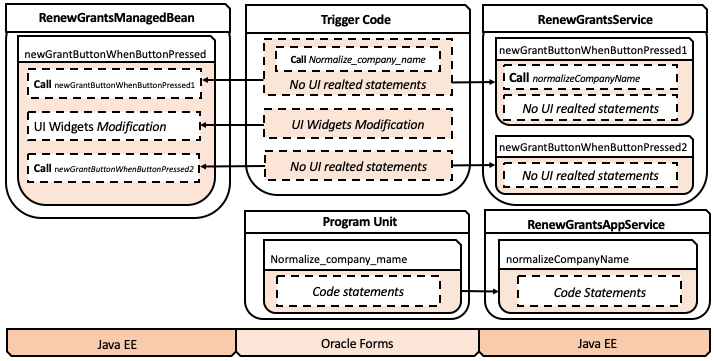}
 \caption{Separating trigger code into managed bean and services.}
 \label{fig:separation-idioms}
\end{figure}

Figure~\ref{fig:separation-idioms} illustrates how the separation strategy
is applied for the trigger example. The first five statements
(lines~2~to~16) do not modify the UI and they are migrated to a
\texttt{newGrantButtonWhenButtonPressed1()} method added to the
\texttt{RenewGrantsService} controller class, and a call to this method is
added to the \texttt{newGrantButtonWhenButtonPressed()} trigger method in the
managed bean class named \texttt{RenewGrantsManagedBean}. This call is
followed by the code that corresponds to the following statement in the
trigger code (lines~18~to~22) because it modifies the UI. Two last
statements do not modify the UI and a new method named
\texttt{newGrantButtonWhenButtonPressed2()} is created in the controller
class. Notice that the program unit that includes the form is migrated to a
method named \texttt{normalize\_company\_name()} that is added to the
\texttt{RenewGrantsAppService} service class.

\paragraph{Sharing variables}
Variables that are present in a trigger
must be shared among the methods that result of its migration: a method in
the managed bean class and one or more methods in the service class. To
achieve this sharing, we have defined a Java map to store the values
associated to each UI attribute. This map contains pairs whose key is the
UI attribute name and the value is the widget value. A managed bean method
pass this map to the invoked service methods. Given the imprecise semantics
of variable scope in PL/SQL, this map is also used to share variable values
among methods.

\paragraph{
Integrating generated code}
Managed Bean class skeletons are generated by the UI Migrator component
developed by Open Canarias. These skeletons contain method headers and the
JSF attributes (declarations and getter/setter methods), but the body is
empty. The implementation of these methods must therefore be completed by
the Code Migrator tool here addressed. This conveys an integration problem
to be tackled in our re-engineering-based solution. The service class with
the methods corresponding to the Program Unit procedures/functions must
also be generated.

\paragraph{Code generated for the trigger example}
Next, we show the Managed Bean and Service classes created for the trigger
example. The class \texttt{RenewGrantsManagedBean} contains the field
\texttt{renewGrantsService} that registers the only Spring service used as
application controller, which is annotated as \texttt{@Autowired}. This
managed bean class also contains the
\texttt{newGrantButtonWhenButtonPressed()} method for the only trigger
defined in the form. This method is organized as explained above (see
Figure~\ref{fig:separation-idioms}). Notice that the
method starts registering values associated to UI attributes in the map
declared to share variables, and ends invoking getter methods declared in
the application controller to obtain the new values calculated.

It calls methods of the \texttt{RenewGrantsService} service in order to
perform database operations and calculate new UI data.

\begin{lstlisting}[language=java]
public class RenewGrantsManagedBean {

  @Autowired private RenewGrantsService renewGrantsService;

  public void newGrantButtonWhenButtonPressed() {
    Map<String, Object> map = new HashMap<String, Object>();
    map.put("year", renewCompanyGrants.getYear());
    map.put("grantCode", renewCompanyGrants.getGrantCode());
    renewGrantsService.newGrantButtonWhenButtonPressed1(map);
    if ((Double)map.get("moneyPaid") >= (Double)map.get("threshold")) {
      setRenewCompanyGrantsGrantRenewedVisible(true); // Generated by UI Migrator component
    } else {
      setRenewCompanyGrantsThresholdNotExceededVisible(true); // Generated by UI Migrator component
    }
    renewGrantsService.newGrantButtonWhenButtonPressed2(map);
    renewCompanyGrants.setThresholdDiference((String)map.get("thresholdDiference")); // Generated by UI Migrator component
    renewCompanyGrants.setTotalAmount((String)map.get("totalAmount")); // Generated by UI Migrator component
  }

  // ... Methods generated by UI Migrator component

}
\end{lstlisting}

The \texttt{RenewGrantsService} class implements the Spring service
corresponding to the controller. This service class contains the
\texttt{newGrantButtonWhenButtonPressed1()} and
\texttt{newGrantButtonWhenButtonPressed2()} methods that correspond to the
two code fragments that do not modify the UI in the PL/SQL trigger. The
first one accesses the database, performs a computation, and updates the
variable map. Then the second one completes the computation of the values
that are lately shown on the user interface. To support the database
access, two methods have been injected: \texttt{writeToDB()} and
\texttt{readFromDB()}. These methods allow to execute arbitrary SQL
statements defined in the source application. The execution is independent
of the quantity of arguments through the JPA technology.

\begin{lstlisting}[language=java]
@Service
public class RenewGrantsService {

  @Autowired private RenewGrantsAppService renewGrantsAppService;
  private EntityManagerFactory emf;

  public void newGrantButtonWhenButtonPressed1(Map<String, Object> map) {
    try {
      String companyName = renewGrantsAppService.normalizeCompanyName(map);
      map.put("moneyPaid", readFromDB("SELECT sum(PAYMENT) FROM GRANTS.GRANTS_PAYMENTS WHERE ...");
      map.put("endowment", readFromDB("SELECT endowment FROM GRANTS.COMPANY_GRANTS ..."));
      map.put("threshold", readFromDB("SELECT threshold FROM GRANTS.COMPANY_GRANTS ..."));
      Double total = ((2 * (Double)map.get("endowment")) - (Double)map.get("moneyPaid"));
      if ((Double)map.get("moneyPaid") >= (Double)map.get("threshold")) {
        writeToDB("UPDATE GRANTS.COMPANY_GRANTS_GRANTED SET state = 'SUSPENDED' WHERE ...");
        writeToDB("INSERT INTO GRANTS.COMPANY_GRANTS_GRANTED (....) VALUES ( ? , ? , ? , ? , ? )", ...);
      }
    } catch (Exception e) {
      message("Database unaccesible")/* TODO: PL/SQL Library Call */;
      throw new FormTriggerFailure();
    }
  }

  public void newGrantButtonWhenButtonPressed2(Map<String, Object> map) {
    Double diference = ((Double)map.get("threshold") - (Double)map.get("moneyPaid"));
    if (diference > 0) {
      map.put("thresholdDiference", diference);
    } else {
      map.put("totalAmount", ((2 * (Double)map.get("endowment")) - (Double)map.get("moneyPaid")));
    }
  }
}
\end{lstlisting}

\paragraph{Code generated for the program unit example}
Finally, the Service class is shown that implements the
\texttt{normalizeCompanyName()} method that is part of the Program Unit.
The generated code is similar to the PL/SQL code. Those functions which
could not be mapped are annotated as manual task to be performed. In the
example, these functions are \texttt{length()} and \texttt{substr()}.

\begin{lstlisting}[language=java]
@Service
public class RenewGrantsAppService {
  public String normalizeCompanyName(Map<String, Object> map) {
    if (length((String)map.get("companyName"))/* TODO: PL/SQL Library Call */ > 256) {
      return substr((String)map.get("companyName"), 1, 256)/* TODO: PL/SQL Library Call */;
    }
    return (String)map.get("companyName");
  }
}
\end{lstlisting}

\section{Overview of the Re-engineering Process\label{sec:Overview}}

\begin{figure}[!ht]
  \centering
  \includegraphics[width=\columnwidth]{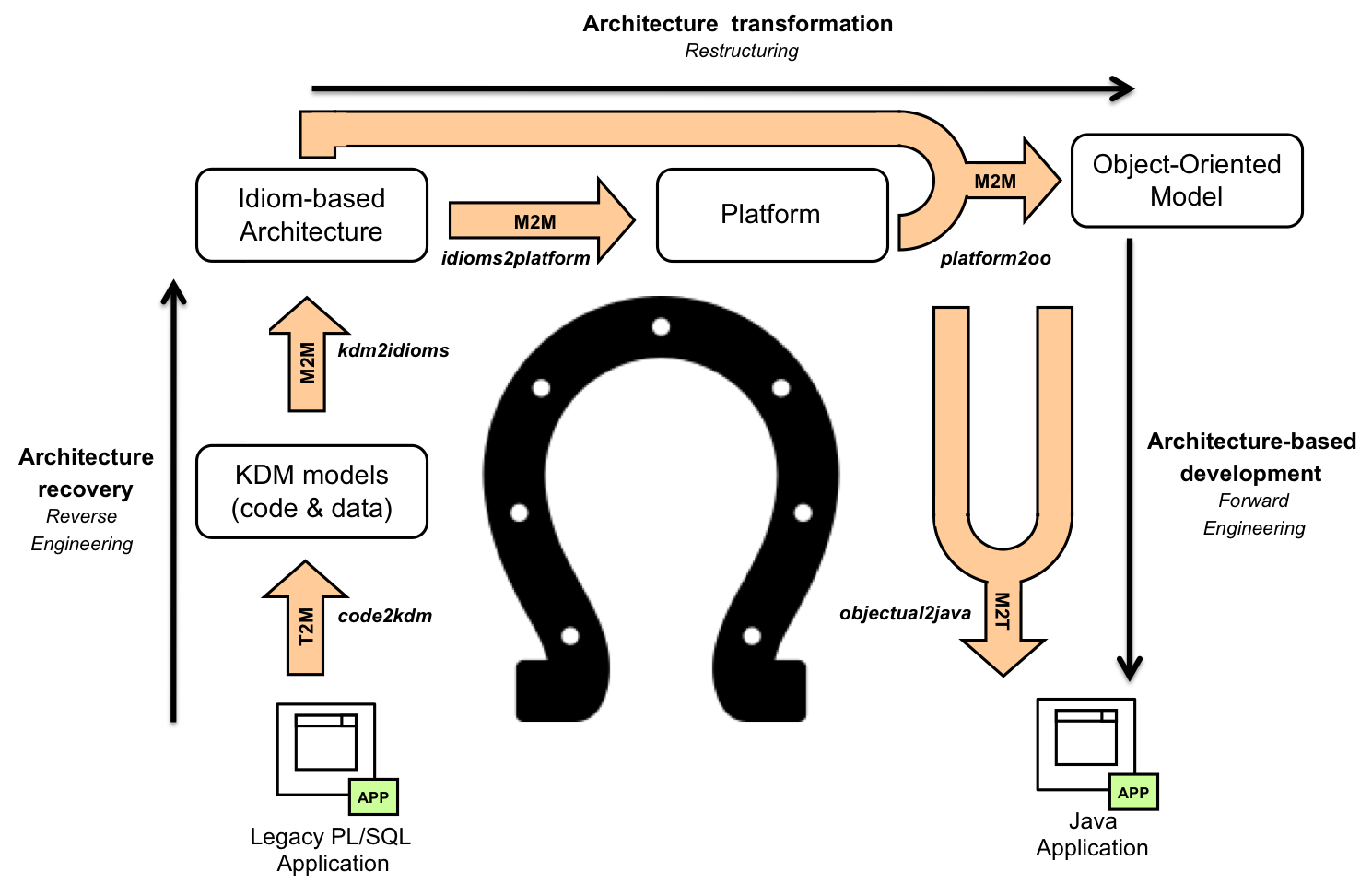}
  \caption{Code re-engineering process in Morpheus.}
  \label{fig:chain-morpheus}
\end{figure}

Figure~\ref{fig:chain-morpheus} shows the horseshoe model for the
model-driven re-engineering process defined for the Code Migrator tool.
A model transformation chain has been implemented to translate
Forms PL/SQL triggers to code of an MVC architecture based on Java
frameworks.

\paragraph{Reverse Engineering}%
In the reverse engineering stage, the behavior of triggers is expressed in
terms of language-independent primitive operations commonly used in RAD
applications (e.g. read data from a database or write data in UI controls),
as proposed in~\cite{wcre11}. Unlike~\cite{wcre11}, we have not considered
RAD idioms (i.e. code patterns) to identify which primitive operations
should a trigger be mapped to. Instead of matching idioms against trigger
code, we analyzed all the statements of the trigger, as the desired level
of legacy code translation has to be of 100\%, as indicated in
Section~\ref{sec:Requirements}. Our solution also differs
from~\cite{wcre11} in other two significant aspects:~(i)~Code and data are
abstracted in form of KDM models, instead of using AST models for the
trigger code; and~(ii)~a different strategy is applied to separate the
legacy code in several concerns in the final application. In addition, we
had to extend the set of primitives considered in~\cite{wcre11} in order to
be able to cover the target~100\% of the PL/SQL code statements.

The reverse engineering stage therefore consists of two steps, as shown in
Figure~\ref{fig:chain-morpheus}. First, PL/SQL code is injected into KDM
models by means of the injector (i.e. the t2m transformation
\emph{code2kdm}) developed by Open Canarias. Next, a m2m transformation
named \emph{kdm2primitives} analyzes the KDM model and creates a model that
conforms to the Primitives metamodel that defines primitive operations. The
stages that generate the KDM models and Primitives models are described in
Sections~\ref{sec:injection} and~\ref{sec:idioms}.

\paragraph{Restructuring} An architectural transformation or restructuring
stage converts a Primitives model into a representation that specifies how
the PL/SQL code is organized in the target platform required by the
company. For this, we have defined the Platform and Object-Oriented
metamodels and a chain of two model-to-model transformations. Both
metamodels are complementary. How the monolithic PL/SQL code is separated
into tiers is represented in a Platform model which has references to
elements of the Object-Oriented model that abstract the code to be
generated. First, a m2m transformation, named \emph{primitives2platform},
generates a Platform model from the Primitives model. This transformation
discovers which classes and methods must be generated. A second m2m
transformation, named \emph{platform2oo}, determines which layer each class
and method belongs to, and obtains a object-oriented representation for
each primitive. This partitioning requires not only information provided by
Primitive models but also accessing to KDM elements that are referenced by
primitives. These two transformations are described in
Sections~\ref{sec:platform} and~\ref{object}.

\paragraph{Forward Engineering}%
Finally, the Platform and Object-Oriented models are used to apply a
forward engineering stage that generates the code that results from the
migration. A m2t transformation named \emph{objectual2java} implements this
stage and generates the Java code that corresponds to the migrated PL/SQL
code. This transformation is described in Section~\ref{code}.

\section{A development process for the Code Migrator tool\label{sec:Development}}

The Code Migrator tool enacts the model-driven reengineering process
described in the previous section. The tool consists of the model
transformation chain that automates the re-engineering approach applied to
move Forms PL/SQL code (triggers and program units) to a Java platform. In
this Section, we will present the software process followed to develop the
tool.

\begin{figure}[!h]
  \centering
  \includegraphics[width=0.95\columnwidth]{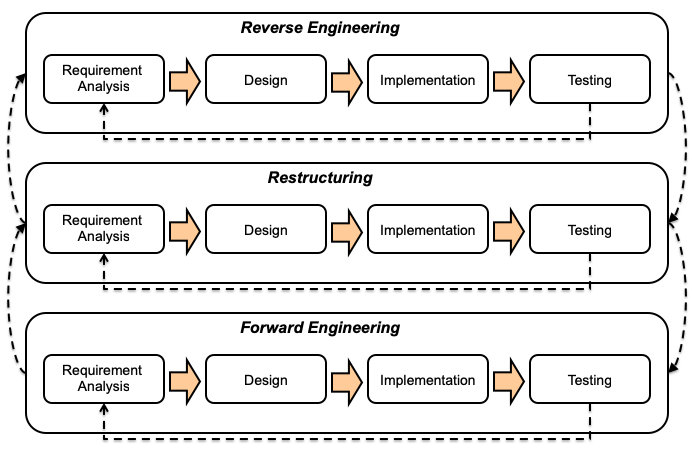}
  \caption{Iterative and incremental life cycle for developing the
    migration tool.}
  \label{fig:lifecycle}
\end{figure}

An iterative and incremental software development life cycle has been
applied, as shown in Figure~\ref{fig:lifecycle}. We established three
iterations, one for each stage of the re-engineering process shown in
Figure~\ref{fig:chain-morpheus}. Each iteration involves requirement
analysis, design, implementation of a model-driven solution for the
corresponding reengineering stage, and testing.

\begin{figure}[!h]
  \centering
  \includegraphics[width=\columnwidth]{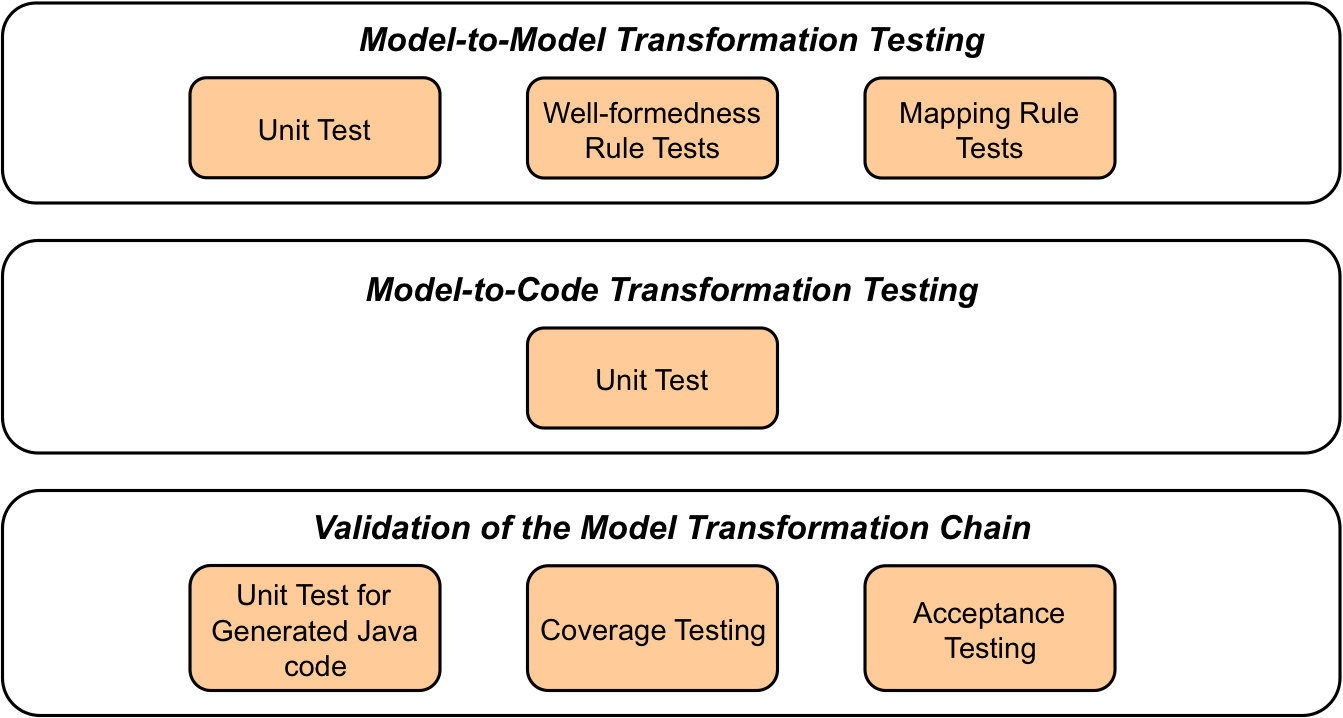}
  \caption{Testing activities in implementing the re-engineering process.}
  \label{fig:testing}
\end{figure}

The greatest effort has been devoted to the implementation and testing
of the model transformations.
The KDM model injector was developed by Open Canarias as indicated in
previous Section. Thus, we tackled the implementation of~3~model-to-model
(m2m) transformations and~1~model-to-text (m2t) transformation. Moreover,
we had to define the target metamodel for each of the m2m transformations.
Testing has been performed for each implemented model transformation, and
once the forward engineering stage was completed, a validation of the
solution was performed, as shown in Figure~\ref{fig:testing}.

Model-to-model transformations are usually complex, specially when the
involved metamodels are large and complex. Therefore, both writing and
testing m2m transformations are challenging activities. Like GPL programs,
m2m transformation definitions must be tested to detect defects in the
implementation. A testing process must be applied to assure that each
transformation in an MDE solution (i.e. a model transformation chain)
properly works.

\subsection{Writing model-to-model transformations in Java}

As part of the re-engineering process we applied to develop the tool, we
factorized the Java code for model-to-model transformations as much as
possible, resulting in the same high-level design for all the
transformations. The code was organized in the following components:

\begin{enumerate}
\item \emph{Iterator}, that traverses the source model.
\item \emph{Analyzer}, that analyzes the current source element to identify
  what target elements must be created to resolve a mapping.
\item \emph{Builder}, that creates and initializes the instances of the
  target model and aggregates these instances into its containing object.
\item \emph{Reference Resolver}, in charge of connecting a created element
  to existing elements according to the references in the target metamodel.
\end{enumerate}

In addition, a particular transformation can have additional components to
implement some specific functionality or utility. The size and
complexity of each component depends on the characteristics of
the transformation.

The Iterator component iterates on the set of root elements of aggregation
hierarchies in the source model, and for each one of them, it traverses all
its aggregation paths. For each visited element, the \emph{Analyzer},
\emph{Builder}, and \emph{Reference Resolver} components collaborate to
create the corresponding target elements. All the m2m transformations will
be explained in detail in Section~\ref{sec:Implementation}. For each
transformation, we shall describe how the components works and how the
testing has been applied.

\subsection{Testing model-to-model transformations}

Model-to-model transformation testing has been extensively addressed in the
literature. Three main challenges are commonly
identified~\cite{baudry2010,fleurey2004,kuster2009}:~(i)~creating a set of
input models (test models) to test the transformation under
study,~(ii)~defining adequacy criteria for checking if test models are
sufficient for testing,~(iii)~checking that the result of a test is
actually the expected model. A discussion on these challenges can be found
in~\cite{baudry2010}, where some emerging approaches to overcome them are
outlined. The source metamodel of a transformation can be used to generate
test models automatically. However, this generation is a complex problem
because it is difficult to assure that a set of models satisfies all the
constraints required for a particular test. Because of this, test models
are usually manually created by using model editors. If there is no editor
specially created for a particular metamodel, model management frameworks,
such as Eclipse/EMF, offer tooling to automatically create a generic editor
for a metamodel. Model comparing tools can be used to check if a
transformation produces the expected result for a given test model.
However, either manually or automatically building expected result models
is again difficult, so m2m transformations are usually validated by
checking a set of constraints on the result model in order to determine if
mappings have been correctly applied.

Some test-driven development approaches have been proposed to implement m2m
transformations. Unit tests have been considered
in~\cite{mantra,jemtte,eunit}, and an incremental process to test model
transformation chains is described in~\cite{kuster2009}, which proposes
four kinds of transformations. 
Some model transformation testing proposals are focused on defining adequately input test models by using: mutation analysis technique~\cite{aranegaMEDBD15}, static analyses~\cite{mottuSTC12}, defining partial test models~\cite{senMTC12} and based-on constraint satisfaction~\cite{senBM09}.
Actually, there is little consensus in how
to effectively test transformations, as it is a very difficult task.
Specially when source and target metamodels are large and complex, such as
those in our project.

\subsection{A test-driven development approach}

\begin{figure}[!h]
  \centering
  \includegraphics[width=0.75\columnwidth]{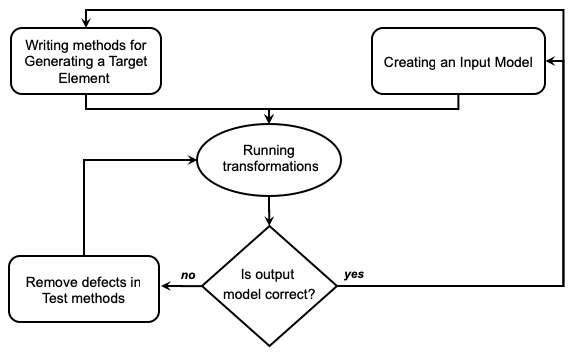}
  \caption{Incremental process for model-to-model transformations.}
  \label{fig:m2m-process}
\end{figure}

On developing the Code Migrator tool, we have defined a test-driven
approach to write model transformations in an incremental way, which is
shown in Figure~\ref{fig:m2m-process}. In each step, the methods that
resolve a particular mapping are written, and a small input model that only
contains instances of the classes involved in the mapping is created. Then,
the transformation is executed for this input model. If errors are found,
the methods are fixed. The output model has always been manually inspected.
Otherwise, a new mapping is considered, and both the methods implementing
it and the test model are created, and the transformation is again
executed. Executed methods are also stored in order to be executed again
later, along with methods for new mappings. 
The process continues until all
the mappings are implemented, and the transformation is completed. 
The decomposition of model transformations in smaller ones 
has been considered in some works as~\cite{sosym15}, where the notion of 
localized transformations is defined.
It should be noted that the implementation of a particular mapping involves
adding methods to the \emph{Analyzer}, \emph{Builder}, and \emph{Reference
  Resolver} components of the model transformation. The \emph{Iterator}
usually does not need to be modified.

Once a m2m transformation is completed following the incremental strategy
commented above, a test is applied to check if generated models satisfy a
set of constraints, which are of two kinds:~(i)~a set of well-formedness
rules for each metamodel, and~(ii)~for each transformation, a set of rules
to check if each mapping has been correctly applied. Both kinds of rules
have been implemented in Java. Checking well-formedness and mapping rules
correspond, respectively, to the integrity and invariant tests considered
in~\cite{kuster2009}.

Regarding validation of the model transformation chain,
we used Form applications of three different sizes as test cases.
This is motivated by the fact that the number of triggers
and program units in a window drastically increases
the complexity to be addressed. Windows with a large number
of associated triggers usually perform a large number of SQL calls,
interface modifications, and program units calls.
Therefore, input models are categorized as small, medium, or large
depending the number of elements per window.
Large input models allow us to validate the algorithm that separate code in
different layers, which cannot be analyzed with small input models.

As shown in Figure~\ref{fig:lifecycle}, three kinds of tests have been
applied. First, \emph{unit tests} were written for the generated code for
each test case. The output code was executed and it was checked that there
are not errors in runtime.
A \emph{coverage testing} is also applied. Metrics about number of
triggers, program units, and code statements have been calculated, as well
as metrics for the equivalent software elements on the target platform.
Then, it has been checked that each previous Forms element is covered on
the Java platform.
Finally, an \emph{acceptance test} is applied, in which each Java event
handler generated is executed to check that it has the same behavior as the
corresponding PL/SQL trigger. The checking is performed manually and
\emph{mocks} have been provided to complete the testing.

It should be noted that no input model had to be created by hand, as an
injector is available to obtain KDM models from PL/SQL code. This saved us
a great effort. We have manually checked that the injected models for the
small, medium, and large test cases include all the PL/SQL statements.

\section{Developing the Trigger Migrator tool\label{sec:Implementation}}

This section explains how the model-driven re-engineering process presented
in Section~\ref{sec:Overview} has been implemented. We will describe each
model transformation involved in the process. The implementation and
testing of each model transformation is explained in detail, except for the
initial t2m transformation (i.e., injection stage) provided by the company.
In our description, we will focus on specifics with respect to the
explanation given in Section~\ref{sec:Development}. In the case of the m2m
transformations, the target metamodels are also described. The trigger
example introduced in Section~\ref{sec:Example} is used to illustrate how
each transformation works.

\subsection{Reverse engineering: KDM injection\label{sec:injection}}

Open Canarias has developed an injector that obtains KDM models from PL/SQL
code.  A detailed explanation of the injector is out of scope of our paper.
In~\cite{canovas2010} a general description of the process to create 
a KDM injector of PL/SQL code. As explained in Section~\ref{sec:KDM}, 
ASTM models can be used to
generate KDM models. An ASTM model represents source code as an AST.
Therefore, a KDM injection process can consist of two stages. Firstly, source
code is transformed in an ASTM model, and then a model-to-model
transformation takes as input the ASTM model and outputs the KDM model.
This process was also followed by the company to build
its injector.

In order to reuse assets in building KDM injectors for different languages,
Open Canarias has also developed the \emph{ModelSET Parser Generator}
framework. This framework provides support for the tasks involved in
implementing a KDM injector:~(i)~generation of the parser that creates a
concrete syntax tree,~(ii)~transforming the concrete syntax tree into an
ASTM model, and~(iii)~transforming ASTM models into KDM models.

This KDM injector developed by Open Canarias is L1 compliant for the Data
package. Therefore, the KDM models extracted--to be reverse
engineered--will be expressed in terms of elements of the \texttt{Code},
\texttt{Action}, and \texttt{Data} packages. The \texttt{Data} package has
not been considered in this paper because we do not migrate the data or the
data structures, just the PL-SQL code, as explained in
Section~\ref{sec:Requirements}.

As shown in Section~\ref{ui-kdm}, the \texttt{UI} package of KDM is very
limited to represent a view structure (layout and kinds of visual
components) and event handling.
Thus, Open Canarias created two families of stereotypes for this package.
The UI structure has been represented by means of stereotypes based on the
USIXML metamodel~\cite{usixml}. On the other hand, UI interaction is
modeled by defining stereotypes based on the IFML metamodel~\cite{ifml}.
Therefore, the injection is not L1 compliant for UI. Moreover, the injector
has not used Micro-KDM, but a family of stereotypes (e.g. SELECT, IF, and
CALL) has been defined to establish the kind of PL-SQL code statement
represented by an \texttt{ActionElement}. The \texttt{kind} and
\texttt{name} attributes of \texttt{ActionElement} are used to indicate the
type and name of the stereotype represented, respectively.

\subsubsection{Application to the Trigger Example}

\begin{figure}[!h]
 \centering
 \includegraphics[width=0.50\columnwidth]{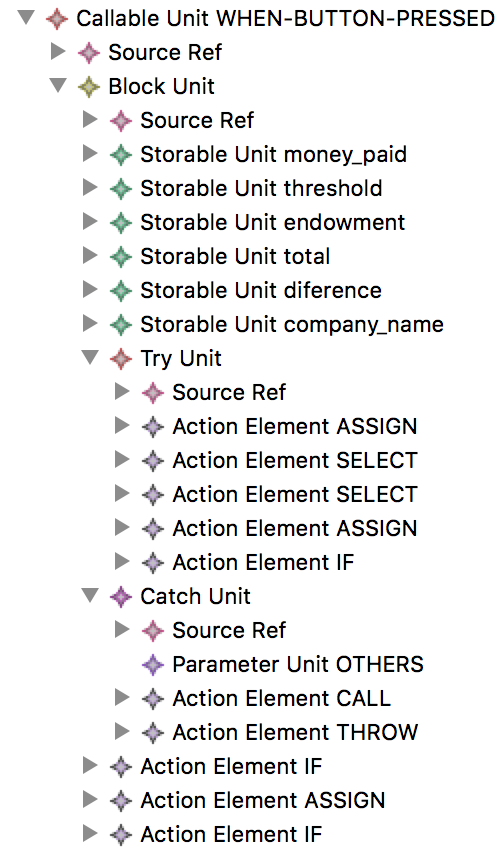}
 \caption{KDM model injected for the trigger example.}
 \label{fig:modelo_ejemplo_kdm}
\end{figure}

Figure~\ref{fig:modelo_ejemplo_kdm} shows the KDM model injected for the
trigger example. An instance of \texttt{BlockUnit} represents the only
trigger of our example. This instance aggregates~(i)~a \texttt{SourceRef}
instance that stores the text of source code,~(ii)~several
\texttt{StorableUnit} instances that represent local variables,
and~(iii)~\texttt{TryUnit} and \texttt{CatchUnit} instances that, in turn,
aggregate the stereotyped \texttt{ActionElement} elements representing code
statements in the trigger code. Depending on the kind of the code
statement, the injector establishes the value to be recorded in the field
\texttt{name} of an \texttt{ActionElement} element as stereotype. In the
example, we can see that \texttt{ActionElement}s with the \texttt{ASSIGN},
\texttt{SELECT}, \texttt{IF}, \texttt{CALL}, and \texttt{THROW} stereotypes
have been created.

This model will be the input of the transformation that completes the
reverse engineering stage by obtaining an Primitives model.

\subsection{Reverse engineering: Generation of the Primitives
  model\label{sec:idioms}}

As indicated in Section~\ref{sec:Overview}, our inference process is based
on the approach presented in a previous work~\cite{wcre11}. Forms triggers
and Program Unit code is represented in terms of primitive operations
commonly used in RAD applications. The use of these primitive operations
provides at higher level of abstraction than PL/SQL statements, and
facilitate the migration to the target platform. Table~\ref{mappings} shows
the primitives defined, which are independent of a particular GPL.

Next, we will first present the Primitives metamodel, and then describe the
\textit{kdm2primitives} transformation that converts injected KDM models
into Primitives models, and finally show the Primitives model obtained for
the trigger example.

\subsubsection{Primitives metamodel}

\begin{figure}[!h]
 \includegraphics[width=\columnwidth]{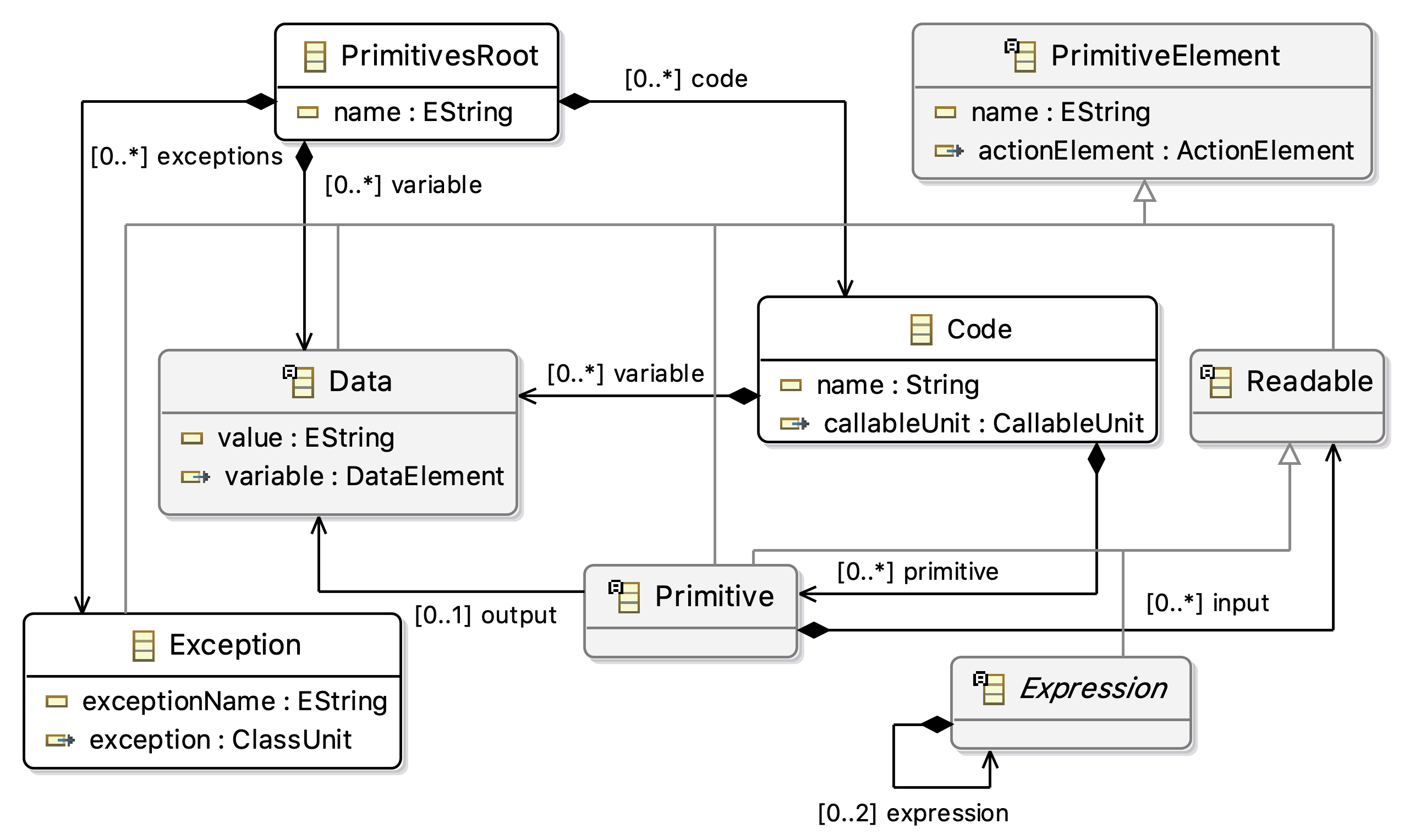}
 \caption{An excerpt of the Primitives metamodel.}
 \label{fig:metaIdioms}
\end{figure}

Figure~\ref{fig:metaIdioms} shows an excerpt of the Primitives metamodel, in
which the hierarchy of classes that represent primitive operations is
omitted. The root class of the metamodel is \texttt{PrimitivesRoot}, which
aggregates three kinds of elements:~(i)~\texttt{Code} that represent the
code of an event handler,~(ii)~a set of \texttt{Variable}s that can have
global or local scope,
and~(iii)~a set of \texttt{Exception}s that can be thrown by the
application.

\texttt{PrimitiveElement} is the root class of the hierarchy of classes
representing all the primitives defined in our metamodel. Primitives are
classified into three categories: \texttt{Variable}, \texttt{Primitive},
and \texttt{Expression}. These classes inherit of \texttt{PrimitiveElement}
and have a reference to the KDM elements from which an primitive was
generated. \texttt{Primitive}, in turn, is the root of the classes that
represent the primitives defined in Table~\ref{mappings}.

A \texttt{Code} is composed of a set of local \texttt{Variable}s, and a set
of \texttt{Primitive}s. A \texttt{Primitive} aggregates a set of zero or
more \texttt{Readable}s as input, and it can reference to a
\texttt{Variable} as output. Each \texttt{Readable} provides an input
value, and the result is stored into a \texttt{Variable}. A
\texttt{Readable} can be a primitive, a variable reference
(\texttt{VariableRef}), or a value returned by a function call
(\texttt{ReturnValue}). An \texttt{Expression} represents a conditional
expression, and can be nested. Below, the classes that represent
expressions are explained.

\begin{figure}[!h]
 \includegraphics[width=1.05\columnwidth]{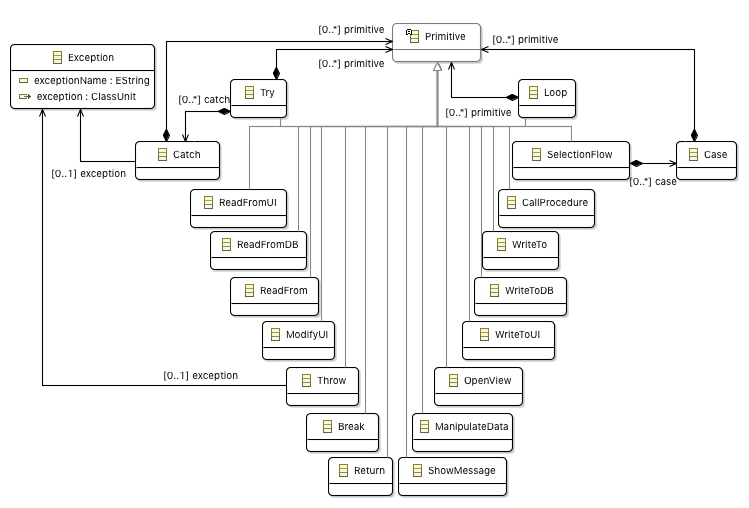}
 \caption{Primitive hierarchy in the Primitives metamodel.}
 \label{fig:primitive}
\end{figure}

Figure~\ref{fig:primitive} shows the hierarchy of \texttt{Primitive}
classes in the Primitives metamodel. \texttt{Loop}, \texttt{SelectionFlow} and
\texttt{Try} represent the three more common primitives.
They are characterized for being composed of other primitives. A
\texttt{Loop} aggregates the \texttt{Expression} that represents the
iteration control condition, and a set of primitives that correspond to the
iteration body. \texttt{SelectionFlow} represents the selection of an
execution flow, either with \emph{if} or \emph{switch} semantics. A
\texttt{SelectionFlow} is composed of a set of \texttt{Case} elements,
which are primitives having associated a condition (\texttt{Expression}
element) and a set of primitives representing the code block to be executed
when the expression is evaluated as \texttt{true}.

For the sake of simplicity, the classes that represent expression elements
have not been shown in the Figures~\ref{fig:metaIdioms}
and~\ref{fig:primitive}, such as~(i)~operators (e.g. \texttt{And},
\texttt{Or}, and \texttt{Less}),~(ii)~the value of a variable
(\texttt{VariableRef}),~(iii)~the return value of a function
(\texttt{ReturnValue}), and~(iv)~kinds of variables: UI widget variable
(\texttt{UIVar}), local variable (\texttt{LocalVar}), global variable
(\texttt{GlobalVar}), and~(v)~constant values (\texttt{Constant}).

\subsubsection{Implementation}

As indicated in Section~\ref{sec:Development}, we have organized the
model-to-model transformations in four components: \emph{Iterator},
\emph{Analyzer}, \emph{Builder}, and \emph{Reference Resolver}. In the case
of the \emph{kdm2primitives} transformation, these four components have
been implemented to perform the following actions:~(i)~to iterate over the
input KDM model,~(ii)~to analyze KDM models to discover
primitives,~(iii)~to create elements that form the Primitives model and
build the aggregation hierarchies, and~(iv)~to establish references between
the primitives newly created and the existing primitives.

For each element in the Primitives model, the \emph{Iterator} invokes the
\emph{Analyzer} which is in charge of discovering primitives in the code.
The \emph{Analyzer} uses the PL/SQL to primitives mappings (see
Table~\ref{mappings}) to decompose code in terms of primitives. Once a new
primitive is identified, the \emph{Analyzer} invokes the \emph{Builder} to
create the corresponding model elements. The KDM elements from which a new
primitive instance will be created are passed as argument in such an
invocation because they provide the information needed to initialize the
elements. Finally, the \emph{Reference Resolver} connects the primitive
created to one or more previously created primitives. It should be noted
that primitives elements maintain a reference to the source KDM elements in
order to keep a traceability to the code that forms a particular usage of a
primitive.

\begin{table}[!ht]
  \caption{Primitives defined for PL/SQL.}
  \label{mappings}
  \centering
\begin{sf}
\begin{scriptsize}
  \begin{tabular}{ll}
    \multicolumn{1}{>{\columncolor[gray]{0.75}}l}{\scriptsize{\textbf{PL/SQL Code}}} &
                                                                                       \multicolumn{1}{>{\columncolor[gray]{0.75}}l}{\textbf{Primitive}/\textbf{Meaning}} \\
    \hline
    \hline

\emph{Variable name}  & \emph{ReadFrom} \\
      & {Read value from a local variable} \\
    \hline
    \emph{Variable name := (Assignment)}  & \emph{WriteTo} \\
     & {Write value to a local variable} \\
    \hline

\emph{UI Variable}  & \emph{ReadFromUI} \\ & {Read value from a UI variable} \\
\hline
\emph{UI Variable := (Assignment)}  & \emph{WriteToUI} \\ & {Write value to a UI variable} \\
\hline

\emph{Select}  & \emph{ReadFromDB} \\ & {Database operation that reads data} \\
\hline
\emph{Insert/Update/Delete}  & \emph{WriteToDB} \\ & {Database operation that insert/update/delete data} \\
\hline

\emph{Operators (+, -, /, *)} & \emph{ManipulateData} \\ & {Arithmetical operation} \\
\hline

\emph{Builtin function e.g. clear\_item} & \emph{ModifyUI} \\ & {Modify an UI property} \\
\hline

\emph{IF / Switch} & \emph{SelectionFlow} \\ & {Execution flow according to conditions} \\
\hline

\emph{While, For, Loop} & \emph{Loop} \\ & {Repeatable code} \\
\hline

\emph{Break} & \emph{Break} \\ & {Ends a loop} \\
\hline

\emph{Procedure/function call} & \emph{CallProcedure} \\ & {Procedure call} \\
\hline

\emph{Return} & \emph{Return} \\ & {Returning value for a funtion/procedure} \\
\hline

\emph{Try} & \emph{Try} \\ & {Code block to execute} \\
\hline

\emph{Throw or Raise} & \emph{Throw} \\ & {Raise an exception} \\
\hline

\emph{ShowMessage} & \emph{Try} \\ & {Open a modal Window} \\
\hline

\emph{OpenView} & \emph{Try} \\ & {Open a Window} \\
\hline
  \end{tabular}
\end{scriptsize}
\end{sf}
\end{table}

\subsubsection{Testing}

The transformation has been implemented following the strategy
explained in Section~\ref{sec:Development}. In this case, the development
process was organized in three phases. In the first phase, we
injected~43~input KDM models for simple triggers that have only one code
statement. With these triggers, we checked expressions, assignments, and
code statements (e.g. IF, CASE, and LOOP) which only includes code blocks
formed for an assignment. In the second phase, we inject~24~KDM models for
triggers containing nested code statements that involves one or more
primitives (e.g. IF or LOOP nested, and TRY-CATCH). In the last phase, we
have taken pieces of code from a real form, in particular the medium-size
form used in the validation of the solution. We have made small changes to
each piece of code to be correctly compiled, and we have used them as input
model.

In each phase, for each KDM model injected, we first write the methods that
implement the corresponding KDM-to-Primitives mapping. Then, the
transformation is executed for the input KDM model, and the output model is
visually analyzed to validate it. For this checking, we have to navigate
through the input KDM model and explore the trigger code. The tree editor
provided by EMF is used to navigate through the models.

KDM models contain a large set of elements, even for simple code fragments.
This in mainly due to the high number of \emph{read} and \emph{write}
elements. For this reason, we decided to use the trigger code to validate
the transformation, additionally to KDM models. Because the high level of
abstraction of Primitives models, it is easy to check if a source code is
correctly represented in terms of primitives.

\subsubsection{Application to the Trigger Example}

\begin{figure}[h]
 \centering
 \includegraphics[width=0.55\columnwidth]{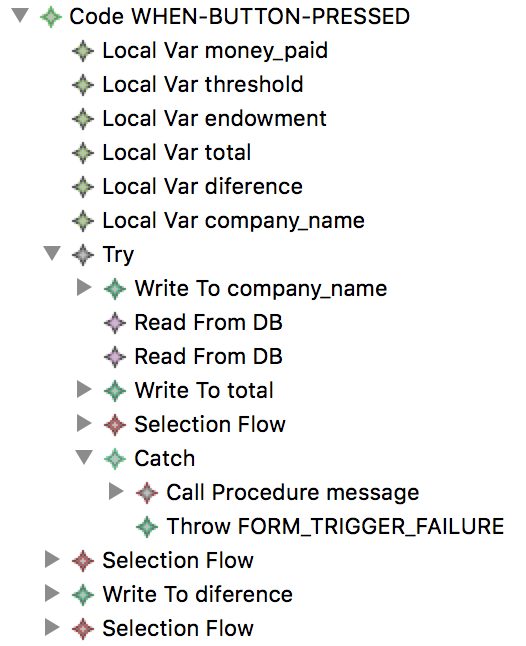}
 \caption{Primitives model example.}
 \label{fig:modelo_ejemplo_idioms}
\end{figure}

Figure~\ref{fig:modelo_ejemplo_idioms} shows the Primitives model obtained
for the example presented in Section ~\ref{code-example}. 
A \texttt{Code} instance has been created which aggregates
the six local variables of the trigger example and the four primitives
identified: a \texttt{Try} (block from line~2 to line~16), two
\texttt{SelectionFlow} (lines~18 and~25), and a \texttt{Write} (write to a
variable in line~24). In turn, the \texttt{Try} element includes the
following primitives: two \texttt{WriteTo} (write to a variable in lines~3
and~7), two \texttt{ReadFromDB} (SELECT operations in lines~4 and~5), a
\texttt{SelectionFlow} (IF statement in line~9), and a \texttt{Catch} that
includes a \texttt{CallProcedure} (call in line~14) and a \texttt{Throw}
(exception triggered in line~15). Primitives elements that are part of
those mentioned above are not shown for the sake of simplicity.

\begin{figure*}[t!]
 \centering
 \includegraphics[width=0.75\linewidth]{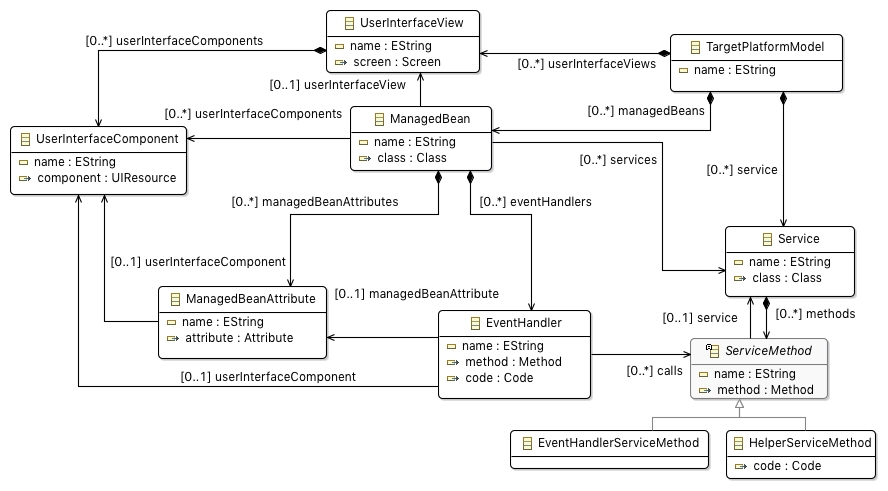}
 \caption{Target Platform Metamodel.}
 \label{fig:metamodelo_plataforma_destino}
\end{figure*}

\subsection{Restructuring: Generation of the Target Platform
  model\label{sec:platform}}

In Section~\ref{sec:Example} we showed how the monolithic code of PL/SQL
trigger and program units is separated into the methods and classes that
correspond to the tiers of the desired MVC architecture: JSF managed beans
and Spring services in our case. This code separation is implemented
through the two model-to-model transformations that restructure the legacy
code in several tiers. First, the \textit{primitives2platform}
transformation generates a \textit{Target Platform} model that represents
the classes and method in \textit{View} and \textit{Controller} tiers. The
model obtained has references to \texttt{Method} objects that are part of a
model that represent the object-oriented code to be generated. This model
is generated in a second m2m transformation, named \textit{platform2oo},
that takes as input the \textit{Platform} model and the \textit{Primitives}
model. Therefore, this second transformation will first map PL/SQL
statements into object-oriented constructs for each method in the
\textit{Platform} model.

We will here describe the first and second transformations in the following
Section. We will begin by presenting the {\em Target Platform} metamodel
that represents the MVC architecture defined in Section~\ref{sec:Example},
more specifically, the JSF and Spring elements involved in the
implementation. It should be noted that although this transformation is
platform-specific, the second one generates a language-independent
object-oriented model.

\subsubsection{Target Platform metamodel\label{platform_mm}}

Figure~\ref{fig:metamodelo_plataforma_destino} shows the Target Platform
metamodel. \texttt{TargetPlatformModel} is the root class of the metamodel.
A \texttt{TargetPlatformModel} aggregates three elements: a
\texttt{Service}, a \texttt{UserInterfaceView}, and a \texttt{ManagedBean}.
A \texttt{Service} represents a \emph{Spring} controller composed of a set
of \texttt{ServiceMethod} methods that are invoked from event handlers or
unit program procedures. A \texttt{UserInterfaceView} represents an
application window and it is made up of a set of
\texttt{UserInterfacesComponent} (UI widgets). A \texttt{ManagedBean}
represents a JSF managed bean. This class is the central element of this
metamodel because it is connected to the rest of the elements. A
\texttt{ManagedBean} aggregates~(i)~the set of event handlers
(\texttt{EventHandler}) of the \texttt{UserInterfaceView} to which it is
associated, and~(ii)~a set of attributes (\texttt{ManagedBeanAttributes})
that denote what UI fields the managed bean manipulates. Moreover, a
\texttt{ManagedBean} references zero or more \texttt{Service}s that
includes the methods \texttt{ServiceMethod} that it calls. It is worth
recalling that \texttt{EventHandler} references a \texttt{Code} element
(i.e. the PL/SQL code of the event handler in form of Primitives) and a
\texttt{Method} element (i.e. the Java code obtained when translating the
PL/SQL code). Finally, two kinds of \texttt{ServiceMethod} have been
defined: \texttt{HelperServiceMethod} for methods created for an operation
enclosed in a program unit, and \texttt{EventHandlerServiceMethod} for
methods created from the trigger code. Like an \texttt{EventHandler}, an
\texttt{EventHandlerServiceMethod} contains references to \texttt{Code} and
\texttt{Method} elements. Instead, a \texttt{HelperServiceMethod} only
references a \texttt{Method}. The need of this distinction will be
evidenced in explaining the implementation of the following two
transformations.

\subsubsection{Implementation}

Next, we explain how managed bean and service classes are generated from
Primitives models. A method is also generated for each PL/SQL trigger,
which is added to a managed bean class, but its body is generated in the
following transformation. When iterating the input Primitives model, it is
required to also navigate over the KDM model because the Primitives
elements do not contain references to the windows of the application. For
this, we take advantage of the existing traceability from \texttt{Code}
elements to the KDM elements that form it. A specific \texttt{KDMNavigator}
class has been created to navigate over the KDM elements, which is used by
the \emph{Iterator} component.

For each \texttt{Code} element in the Primitives model, the
\texttt{CallableUnit} KDM element that references it is accessed. From this
element, a bottom-up navigation is performed from the UI component to which
the trigger is associated (\texttt{UIResource} KDM element) to the parent
application window (\texttt{Screen} KDM element). Then, a
\texttt{UserInterfaceComponent} is instantiated, which references to the
\texttt{UIResource} that is source of the event handled by the trigger.
When instantiating the first \texttt{UserInterfaceComponent}, a
\texttt{UserInterfaceView}, \texttt{ManagedBean}, and \texttt{Service}
object are then created. Each created \texttt{UserInterfaceComponent} is
added to the \texttt{UserInterfaceView} and is referenced from the
\texttt{ManagedBean}. Whenever a \texttt{UserInterfaceComponent} is
instantiated, an \texttt{EventHandler} and a \texttt{ManagedBeanAttribute}
are also created, which references it. The attribute \texttt{code} of
\texttt{EventHandler} is initialized with a reference to the current
\texttt{Code}. As indicated above, each \texttt{EventHandler} object has a
reference to the \texttt{Method} object that represent its object-oriented
code. These references will be void when the \textit{primitives2platform}
transformation is completed, and they will be initialized during the
execution of the \emph{platform2oo} that has as input the Platform model.

When the visited \texttt{Code} element refers to a Program Unit, only a
\texttt{Service} is created (not a \texttt{ManagedBean}). A Program Unit is
identified if the \texttt{CallableUnit} accessed from the \texttt{Code}
element does not contain a reference to an \texttt{UIResource} but it
contains a \texttt{CodeFragment} stereotype.

Finally, it should be noted that \texttt{ServiceMethod}s for a
\texttt{Service} generated for a \texttt{UserInterfaceComponent} can not be
created in this transformation, because they are created depending on the
structure of the trigger code. The analysis of this structure is performed
in the following transformation, when primitives are translated into
object-oriented constructs. Instead, only \texttt{ServiceMethod}s for a
\texttt{Service} that are originated from a Program Unit are generated.
This explains the existence of two kinds of \texttt{ServiceMethod}s.

\subsubsection{Testing}

The transformation has been implemented in two phases. First, we have
injected a KDM model for different program units in an input form. We
checked that the target platform had a service and as many methods as
program units. Second, we created more complex forms by adding different
kinds of widgets, nesting windows, and having more than one window, in all
the cases having a trigger associated to each visual component contained in
the form. Then, we checked that~(i)~the platform model generated included a
\texttt{UserInterfaceView} for each window and a
\texttt{UserInterfaceComponent} for each widget,~(ii)~a
\texttt{ManagedBean} and a \texttt{Service} for each window, and~(iii)~a
\texttt{ManagedBeanAttribute} and \texttt{EventHandler} for each visual
component. In each phase we applied the incremental development strategy
discussed in Section~\ref{sec:Development}.

\begin{figure*}[!t]
 \centering
 \includegraphics[width=0.85\linewidth]{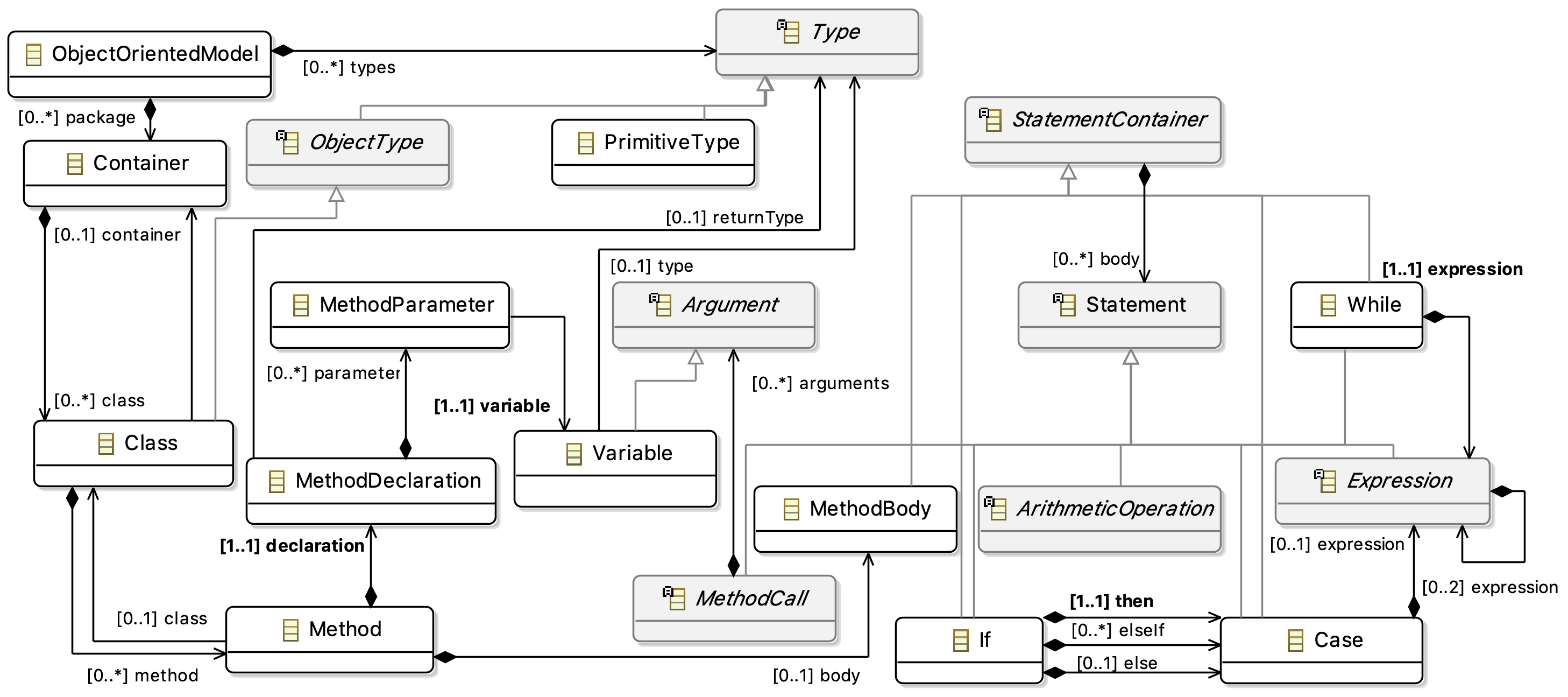}
 \caption{An Object-Oriented metamodel based on the Java metamodel of
   Modisco.}
 \label{fig:metamodelo_orientados_objetos}
\end{figure*}

\begin{figure}[!t]
 \centering
 \includegraphics[width=0.80\columnwidth]{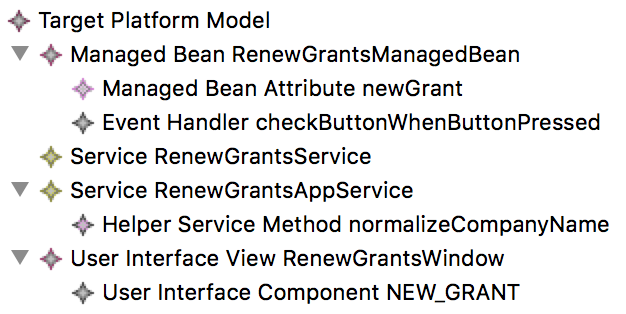}
 \caption{Target Platform Model example.}
 \label{fig:modelo_ejemplo_plataforma}
\end{figure}

\subsubsection{Application to the Trigger Example}

Figure~\ref{fig:modelo_ejemplo_plataforma} shows the main elements of the
target platform model generated for the example:~(i)~one
\texttt{ManagedBean},~(ii)~one \texttt{UserInterfaceView}, and~(iii)~two
\texttt{Service}s. The \texttt{ManagedBean} contains one
\texttt{ManagedBeanAttribute} that refers to the
\texttt{UserInterfaceComponent} representing the button and an
\texttt{EventHandler} that contains a reference to the event handler
method. The \texttt{GrantsService} has not
\texttt{EventHandlerServiceMethod}s because them are generated in the
following transformation, while the \texttt{GrantsAppService} service
contains a \texttt{HelperServiceMethod} that corresponds to the function
\texttt{normalizeCompanyName} that is part of the program unit. This
service method has a null value for the attribute \texttt{method} as the
\texttt{Method} instance will be created in the following transformation as
explained above. Finally, the \texttt{UserInterfaceView} element aggregates
an \texttt{UserInterfaceComponent} element which correspond to the only
widget that it contains (i.e. a \emph{button}). Note that only those visual
components that have a \emph{trigger} associated are aggregated.

\subsection{Restructuring: Generation of the Object model\label{object}}
\label{sec:oo}

We shall describe the \emph{platform2oo} transformation that, taking the
Primitives and target platform models as input, is able to generate the
Object-Oriented model. While the target platform model specifies which
classes and methods must be generated for each tier of the MVC
architecture, the Primitives model provides an abstract representation of
the legacy code. Therefore, the target Platform model directs the
transformation, and the Primitives model is needed to generate the
object-oriented representation of the code to be included in each method.
First of all, we will present the Object-Oriented metamodel.

\subsubsection{Object-Oriented metamodel}

Figure~\ref{fig:metamodelo_orientados_objetos} shows the Object-Oriented
metamodel. Reusing a model is usually a better option than building one
from scratch. In our case, several object-oriented metamodels are
available, such as FAMIX~\cite{famix} and the Modisco's Java
metamodel~\cite{modisco}. We reused the latter because we were more
familiar with it, but it was modified in order to have a
language-independent metamodel. Next, we will comment the main elements of
this metamodel.

An \emph{Object-Oriented model} consists of a set of \texttt{Type}s and
\texttt{Module}s. A \texttt{Class} contains a set of \texttt{Attribute}s
and \texttt{Method}s. A \texttt{Method} is composed of a header
(\texttt{MethodDeclaration}), a body (\texttt{MethodBody}), and local
variables (\texttt{Variable}). The code statements of a method body are
represented as a set of \texttt{Statement}s. \texttt{Statement} is the root
of a hierarchy representing the kinds of code statements.
\texttt{StatementContainer} inherits from \texttt{Statement} and, in turn,
is the root of statements that can include a code block as: conditional
(\texttt{If} and \texttt{Switch}) loops (\texttt{For}, \texttt{While}, and
\texttt{ForEach}), and exception handlers (\texttt{Try} and
\texttt{Catch}). Other kinds of code statements are expressions
(\texttt{Expression}), assignments (\texttt{VariableAssign}) method calls
(\texttt{MethodCall}), and variable declarations
(\texttt{VariableDeclaration}).

\subsubsection{Implementation}

The \emph{Iterator} component traverses the \texttt{ManagedBean} and
\texttt{Service} elements contained in the Platform model. For each
\texttt{ManagedBean} and \texttt{Service}, the \emph{Builder} component
creates a \texttt{Class} whose name is formed by a window identifier
followed by ``ManagedBean'' or ``Service'', respectively. The window
identifier results of concatenating the window name with the value of a
counter used to accumulate the number of windows found in a form. Once
these classes are created, the transformation must generate
\texttt{Method}s and \texttt{Attribute}s for each of these classes. This
generation is done differently for \texttt{ManagedBean} and
\texttt{Service} as explained below. Note that the process performed
involved the components \emph{Analyzer}, \emph{Builder}, and
\emph{Reference Resolver}.

\paragraph{Services processing for program units} As explained above, the
Platform model contains a \texttt{Service} that aggregates a
\texttt{HelperServiceMethod} for each operation that is part of a program
unit. For these services, a \texttt{Method} is generated for each
\texttt{HelperServiceMethod}. The \texttt{code} attribute of a service
method is used to access the set of primitives that represent the
corresponding program unit operation. These primitives are translated into
the object-oriented statements that form the created method body. In this
way, we have a \texttt{Service} class with a method for each operation
included in a program unit.

\paragraph{Event handlers processing}
Here we describe in detail the implementation of the code separation
strategy of PL/SQL triggers outlined in section~\ref{Code-separation} and
illustrated in Figure~\ref{fig:separation-idioms}. Note that legacy code
has already been analyzed to identify the managed bean classes and event
handler methods to be generated (i.e. the target platform model).
Algorithm~\ref{alg:separation} depicts the processing of the event handlers
of a managed bean.

\begin{algorithm}
\caption{Separation of Managed bean and Service code.}
\label{alg:separation}
\begin{algorithmic}[1]
\scriptsize
\For {$eventHandler \in managedBean.eventHandlers$}
        \State{$processCode(eventHandler)$}
\EndFor
\\
\Procedure{$processCode$}{$eventHandler$}

\State{$eventHandler.method \gets createMethod()$}
\State{$method \gets eventHandler.method$}
\State{$code \gets eventHandler.code$}
\\
\State{$serviceMethod \gets createServiceMethod()$}
\State{$method.createServiceCall(serviceMethod)$}
\For {$primitive \in code.primitives$}
        \State{$moveToUi \gets processPrimitive(primitive, serviceMethod.statements)$}
        \If {$moveToUi$}
                \State{$serviceMethod.moveLastStatementToMethod(method)$}
                \State{$serviceMethod \gets createServiceMethod()$}
                \State{$method.createServiceCall(serviceMethod)$}
        \State{$method.moveVariablesToMethod(serviceMethod)$}
        \EndIf
\EndFor
\If {$service.statement.isEmpty()$}
        \State{$method.removeLastServiceCall()$}
         \State{$delete serviceMethod$}
\EndIf

\EndProcedure
\\
\Function{$processPrimitive$}{$primitive, statements$}
        \State{$OOstatement \gets doMapping(primitive)$}
                                \State{$statements.add(OOstatement)$}
                                \State{$moveUI \gets false$}
        \If {$isComplexPrimitives(primitive)$}
                \For {$nestedPrimitives \in primitive.primitives$}
                        \State {$moveToUI \lor processPrimitive(nestedPrimitives, OOstatement.statements)$}
                \EndFor
        \ElsIf {$isModifyUIPrimitive(primitive)$}
                \State{$moveUI \gets true$}
        \EndIf
        \State{\Return{$moveUI$}}
\EndFunction
\end{algorithmic}
\end{algorithm}
\normalsize

The set of \texttt{EventHandler}s that aggregates a \texttt{ManagedBean}
is traversed in order to apply the code processing (line~1) on each of
them. First of all, a new \texttt{Method} is created containing the code
statements of the event handler code in the managed bean (line~6).
Given the set of primitives describing a trigger (the \texttt{Code} instance
referenced by the \texttt{code} attribute of the event handler) and the
created method (the \texttt{method} attribute),
the algorithm first creates another method that is added to the service
class (line~10), and a method call that is added to the managed bean method
(line~11). When the service method is created, a
\texttt{EventHandlerServiceMethod} is also created and aggregated to the
\texttt{Service} instance of the Platform model. Then, the primitives that
form an event handler are traversed in order to separate code manipulating
view elements from the rest of the code (line~13), as illustrated in
Figure~\ref{fig:separation-idioms}.

Each primitive is processed to check whether it is a \texttt{ModifyUI}
primitive (line~27). This processing first translates each primitive into a
set of object-oriented statements, which are added to a list of statements
(line~29), and after it is checked if the current primitive is a
\texttt{ModifyUI} (line~35) or either it has other nested primitives. When
nested primitives are found, they are recursively processed (line~33).
Whenever a \texttt{ModifyUI} is found four actions are sequentially
executed: the last primitives processed are translated into object-oriented
statements and moved to the current service method (line~15), a new service
method is created (line~16), a method call to this method is created and
added to the managed bean method (line~17), and shared variables are also
moved to the new service method (line~18).

\paragraph{Sharing variables among generated methods}
The problem of identifying what variables are being shared among methods
that result of code separation is not trivial, as indicated in
Section~\ref{sec:background}. The previous algorithm also declares and
shares variables by means of storing data about accesses to variables. This
information allows us to discover those variables which are being used in
several methods and therefore which of them are being shared. When the
algorithm detects one read/write (in the \texttt{processPrimitive}
function), the primitive is processed as a local variable access and all
the information is stored: the variable, the variable access, and the
method where the variable is being used. The method can be a service or a
managed bean. Special consideration must be taken into account when the
variable access is done inside a complex primitive that contains a
\texttt{ModifyUI} primitive. Then, the code of the complex primitive is
moved from the service to the managed bean. In addition, the method
associated to the variable access is updated.

Once all the primitives of the code trigger have been processed, the
variable accesses contained in a method are compared to all the variable
accesses of the rest of methods. When a variable access is present in two
or more methods, then it is required to change the local accesses of the
variable by the use of a map as discussed in Section~\ref{sec:Example}. The
map is in charge of sharing variable values between different methods by
sharing the same map instance among all of them. When the variable is only
accessed by one method, then a local variable declaration inside the method
is produced. Because in PL/SQL code variables are declared at the beginning
of the trigger procedure, the algorithm requires to calculate the adequate
place where to introduce the variable declaration.

\paragraph{Declaring variables}
In the following, we describe the part of the algorithm for discovering
the code blocks where a variable declaration must be introduced (see
Algorithm~\ref{alg:declaring}). The first variable access is set as the
variable declaration. Then, the algorithm iterates over all the variable
accesses, and it compares each one to the initial variable declaration to
check in what code block the variable declaration must be implemented.
For each pair of initial variable declaration and variable access, the
block where both are included is compared. If both blocks are the same,
the first variable access is chosen as the variable declaration (line~4).
For instance, a source code as \texttt{age~:=~18} will produce the next
target code: \texttt{int age~=~18;}. If blocks are different, then it is
checked if one block is containing the other one. In that case, the outer
block will contain the variable declaration (lines~6 and~8). If blocks are
different and there is not a containment relation, then the first block
containing both blocks is chosen to implement the variable declaration
(line~10). For example, the next source code \texttt{if (age >{}= 18) adult
  = true; else adult = false} will determine that declaration must be in
the block immediately containing the block \texttt{if (boolean) \ldots{}
  else \ldots}. Finally, if the first variable access corresponds to a
read, a code comment is introduced in order to inform that a variable is
being used and could not have been initialized. It is worth recalling that
the Java compiler initializes to \texttt{0} all \texttt{int} variables not
initialized explicitly in the code. However, in PL/SQL the meaning of an
initialization absence would be the assignment of a \textit{NULL} as
initial value, whichever the variable type was.

\begin{algorithm}
\caption{Declaring variables in the right code block.}
\label{alg:declaring}
\begin{algorithmic}[1]
\scriptsize

\State{$variableDeclaration \gets firstVariableAccess$}
\For {$access \in restOfVariableAccesses$}
        \State{$declarationBlock \gets empty$}
        \If {$variableDeclaration.block == access.block$}
                \State {$declarationBlock \gets variableDeclaration.block$}
        \ElsIf {$variableDeclaration.block \not= access.block \land \newline
                        access.block \in variableDeclaration.block$}
                \State{$declarationBlock \gets variableDeclaration.block$}
        \ElsIf {$variableDeclaration.block != access.block \land \newline
                        variableDeclaration.block \in access.block$}
                \State{$declarationBlock \gets access.block$}
        \Else
                \State{$declarationBlock \gets allBlocks.selectFirst( block |  \newline
                block.contains(variableDeclaration.block) \land \newline
                block.contains(access.block))$}
        \EndIf

        \If {$isOnlyRead(firstVariableAccess)$}
                \State{$declarationBlock.addComment("// Variable not explicitly initialized")$}
        \EndIf

\EndFor

\end{algorithmic}
\end{algorithm}
\normalsize

\subsubsection{Testing}

The part of the transformation that concerns to the procedural-to-object
translation has been incrementally built by reusing the primitives models
that were generated during the implementation of the \emph{kdm2primitives}
transformation. Recall that we injected~67~KDM models which covered the set
of defined primitives. Therefore, we have checked that the Object-Oriented
model generated for each of the~67 Primitives model previously generated is
correct. For this, we have manually explored each Primitives model and
checked that the expected object-oriented statements have been generated
for each primitive. Each of the object-oriented models has been easily
validated, because the mapping between primitives and object-oriented
elements is simple. We have also performed this process for the Primitives
models generated from pieces of code taken from the medium-size form.

Regarding the implementation of the separation of code, the transformation
has been validated through the code generated by the m2t transformation
explained below. The direct correspondence between the object-oriented
model and the generated Java code justifies this decision. We found that
the manual validation of the code separation was easier on Java code
editors than using modeling editors. In addition, the Java compiler checks
automatically if a variable was declared in the right code block.

\subsubsection{Application to the Trigger Example}

\begin{figure}[!ht]
 \centering
 \includegraphics[width=0.60\columnwidth]{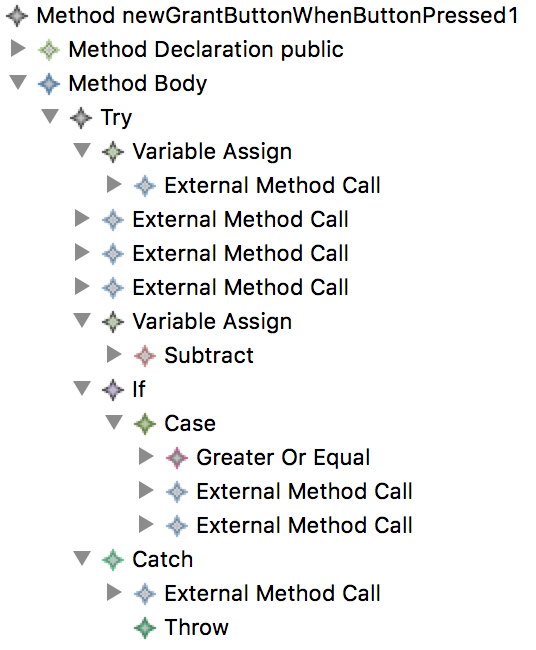}
 \caption{An excerpt of the Object-Oriented model for the example.}
 \label{fig:modelo_ejemplo_orientado_objetos}
\end{figure}

Figure~\ref{fig:modelo_ejemplo_orientado_objetos} shows an excerpt of the
Object-Oriented model obtained for the example. Specifically, the method
\texttt{newGrantButtonWhenButtonPressed1} included in the class
\texttt{RenewGrantsService}. The method contains the elements that
corresponds to the code shown in Section~\ref{sec:Example}. It starts with
a \texttt{Try} and several calls to methods (in this case, first the call
to \texttt{normalizeCompanyName}, then the different \texttt{readFromDB}).
Then, after the assignment of the variable, the \texttt{If}, and finally
the corresponding \texttt{Catch}, with its body.

\subsection{Forward engineering: Code generation\label{code}}

A simple m2t transformation named \emph{objectual2java} has been
implemented to generate the final application code from the Object-Oriented
model. This transformation has been implemented in Acceleo~\cite{acceleo}.

The transformation is straightforward. However, it had to be integrated
with the \emph{UI2Java} m2t transformation created by Open Canarias as part
of the UI Migration tool in order to generate code for the user interfaces.
\texttt{ManagedBean} class skeletons are generated by the \emph{UI2Java}
transformation. Therefore, our \emph{objectual2java} transformation must
generate all the code for service classes and code for the method body of
the \texttt{ManagedBean} classes. The strategy agreed with the company to
integrate both transformations is the following: \emph{UI2Java} templates
invoke to \emph{objectual2java} templates passing a \texttt{CallableUnit}
KDM element as argument. Then, we traverse the Platform model to find the
\texttt{EventHandler} whose \texttt{Code} references this
\texttt{CallableUnit}, and we generate Java code for the \texttt{Method}
element referenced from that \texttt{EventHandler}. Therefore, the
\emph{objectual2java} transformation also has the
\emph{Object-Oriented} and \emph{Platform} models as input.

\subsubsection{Testing}

Because the object-oriented models are representations that are very close
to the Java code, the Acceleo templates are simple and they are therefore
easy to test.
We have followed the same strategy used to validate Object-oriented models.
That is, we have used the models object-oriented generated from the~67~KDM
models injected from simple triggers. In fact, object-oriented models and
Java code are validated at the same time. Therefore, this model-to-text
transformation is written the same as the \emph{platform2oo} transformation
is written. Generated code was compiled for syntactic and semantic
validation. Moreover, a manual validation was carried out to check if
indentation was correct for each line of code, and code statements follow
the correct order.

Once the \emph{objectual2java} transformation has been completed, we
checked that the \texttt{ManagedBean} and \texttt{Service} classes have
been correctly generated, which implies to perform five kinds of
checks:~(i)~a managed bean class and a service class has been generated for
each trigger in a form;~(ii)~a service class has been created for all the
program units of a form, and~(iii)~the methods of each class have been
generated, and they have the expected code. This validation has been
carried out for the three forms presented in the following section.

\subsubsection{Application to the Trigger Example}

The code generated for the trigger example was shown in
Section~\ref{sec:Example}. As can be observed there, the following
artifacts were generated:

\begin{itemize}
\item A managed bean class \texttt{RenewGrantsManagedBean} with a event
  handler method \texttt{newGrantButtonWhenButtonPressed()}.
\item A service class \texttt{RenewGrantsService} with the two methods that
  are invoked from the event handler method:
  \texttt{newGrantButtonWhenButtonPressed1()} and
  \texttt{newGrantButtonWhenButtonPressed2()}. The first one corresponds to
  the excerpt of Object-oriented model shown in
  Figure~\ref{fig:modelo_ejemplo_orientado_objetos}.
\item A service class \texttt{RenewGrantsAppService} for the program unit
  included in the form, which contains the method
  \texttt{normalizeCompanyName()}.
\end{itemize}

It is worth noting that the service class \texttt{RenewGrantsService} has
two methods because the trigger example has only a \texttt{ModifyUI}
primitive operation located between two code fragments that correspond to
other primitives. The code of this \texttt{ModifyUI} primitive is shown
below.

\begin{lstlisting}[language=SQL,morekeywords={END,IF}]
IF money_paid >= threshold THEN
  SET_ITEM_PROPERTY('RENEW_COMPANY_GRANTS.GRANT_RENEWED', visible, property_true);
ELSE
  SET_ITEM_PROPERTY('RENEW_COMPANY_GRANTS.THRESHOLD_NOT_EXCEEDED', visible, property_true);
END IF;
\end{lstlisting}

The Java code generated for this \texttt{ModifyUI} primitive would be:

\begin{lstlisting}[language=java]
if ((Double)map.get("moneyPaid") >= (Double)map.get("threshold")) {
  setRenewCompanyGrantsGrantRenewedVisible(true);
} else {
  setRenewCompanyGrantsThresholdNotExceededVisible(true);
}
\end{lstlisting}
\section{Validation of the tool\label{sec:Validation}}

This Section is devoted to describe the validation of the final code
generated by the tool (i.e. the output of the transformation chain that
implements the re-engineering process). We first establish the scope of the
validation by providing a definition of all the tests involved. Then, a
description of the applied instrumentation and methodology is given.
Finally, we present and comment the result of the validation. In addition,
the threats to validity that should be considered in order to accept the
result are identified.

\subsection{Definition}

Validating a model transformation chain entails to test each transformation
as well as to check whether the final software artifacts generated meet the
initial requirements. In the case of a code migration, these requirements
establish how the legacy code must be translated into target code.

In the previous Section, we have explained the black-box testing approach
applied to validate each of the transformations in the chain. Here, we will
explain the tests performed to validate that the tool produces the expected
code. As indicated in Section~\ref{sec:Development}, three kinds of
black-box tests have been considered to verify how accurate the obtained
results are, and how they conform to source-target mapping.

\begin{itemize}

\item \emph{Unit tests for generated Java code}. Unit tests have been
  written and executed for all the generated classes.
\item \emph{Migration coverage tests}. Some software metrics have been used
  to measure the migration coverage. Coverage tests are executed to know
  how much code constructs in the legacy application have been migrated.
  These measurements are useful to verify that all the legacy code has been
  processed during the migration process.
\item \emph{Acceptance tests}. Functional requirements have been finally
  validated by means of acceptance tests. For this, the company has checked
  that legacy PL/SQL code and generated Java code have the same behavior
  for the three forms described later.
\end{itemize}

\subsection{Instrumentation}

JUnit has been used to write and run the unit tests for the generated
Managed Bean and service classes. Additionally, we have used a Java
compiler to check the syntax and some semantic aspects such as method
invocations.

The coverage testing has involved the use of tools for software inspection.
Concretely, we have used the ClearSQL7 and Eclipse Metrics~3 tools for
obtaining metrics from Oracle Forms applications and Java code,
respectively. The metrics measure the number of triggers, program units,
and SQL code statements for the source application, and the number of
methods and SQL code statements for the target application. It is worth
recalling that triggers and program units are migrated to methods in the
target platform. In order to validate the intermediate models, we have
implemented dedicated Java/EMF programs that calculate the metrics
indicated above for the generated Ecore models.

Acceptance tests have been manually performed by testers from the company.
They have executed the legacy and target applications to check whether UI
event handlers have the same behavior.

The three kinds of testing have been carried on three legacy forms provided
by the company, which have different size. As indicated in
Section~\ref{sec:Development}, the size of these three forms is categorized
as \emph{small}, \emph{medium}, and \emph{large} depending on the number of
elements per window, as shown in Table~\ref{sizeforms}. This table shows
the number of triggers, program units and lines of code in the three
forms.

\begin{table}[!ht]
  \caption{Triggers, Program Units and LoC of the tested
    forms.\label{sizeforms}}
\begin{sf}
\begin{scriptsize}
\centering
\begin{tabular}{p{1.1cm}rrrrr}

 \multicolumn{1}{>{\columncolor[gray]{0.90}}l}{{\textbf{Forms}}} &
 \multicolumn{1}{>{\columncolor[gray]{0.75}}l}{{\textbf{\#triggers}}} &
 \multicolumn{1}{>{\columncolor[gray]{0.90}}l}{{\textbf{triggers LoC}}}&
 \multicolumn{1}{>{\columncolor[gray]{0.75}}l}{{\textbf{\#units}}} &
 \multicolumn{1}{>{\columncolor[gray]{0.90}}l}{{\textbf{units LoC}}} &
 \multicolumn{1}{>{\columncolor[gray]{0.75}}l}{{\textbf{Total LoC}}} \\
  \hline
  \hline
        small & 5 & 807 & 25 & 442 & 1249 \\
        medium & 42 & 2206 & 30 & 1302  & 3508 \\
        large & 251 & 6563 & 68 & 4617 & 11180
\end{tabular}
\end{scriptsize}
\end{sf}
\end{table}

\subsection{Methodology}

At the end of the implementation of each model transformation, we measure
the metrics defined for coverage testing. They are calculated by executing
the Java program that corresponds to the target metamodel. In addition, we
have also manually calculated the metrics by inspecting the target model
with the EMF tree editor. In the case of the Primitives model, we make sure
that the model elements are also present in the KDM model and the source
code. Platform models have been validated by checking if each UI widget,
event handler, window, trigger and program unit has been correctly
injected. Object-oriented models are validated in a similar way as
Primitives models. Moreover, these models are indirectly validated through
testing on the generated code, because there are a direct correspondence
between them.

Once the model transformation chain was developed, unit tests were first
tackled. Before writing the unit tests, the generated code was modified to
include some mocks to simulate the built-in functions and the data access
methods invoked. Then, unit tests were written for the managed bean and the
service classes. The unit tests were performed following a bottom-up
strategy according to the dependencies among classes. First, the unit tests
for the unit programs service class were performed, then for the service
class implementing the data access and the application logic, and finally
for the managed bean class that includes the event handler methods that
invoke the methods of the service classes.

After running the unit tests, we measured the metrics that we mentioned
before. The results are shown in Table~\ref{resultados}, and are commented
below.

We performed integration tests on the services, without considering the
calling methods to mocks. Then we proceed in a similar way on the managed
beans.

Finally, company testers executed the generated code interacting with the
generated JSF views for the three tested forms. As explained in
Section~\ref{code}, the methods generated by the \textit{objectual2java}
m2t transformation, are invoked from the templates generated by the UI
Migration tool. Therefore, the managed bean and service classes are
automatically integrated in the views classes generated by the UI Migration
tool. Prior to perform the acceptance tests, company members developed the
integration tests for the data access services to validate the integration
of the managed bean classes with the service classes. As a result of the
acceptance tests, some errors were reported, for example, identifiers that
were declared as local variable and parameter in the same method, missing
import code statements, the \texttt{i++} code statement was generated as
\texttt{i=+}, or some primitive type variables with a NULL value. These
errors were fixed in the model transformation chain.

\subsection{Results}

The results obtained for the three tested forms are shown in
Table~\ref{resultados}. The column \emph{\#elements} indicates the measured
metrics. As depicted in the table, the metrics count the number of legacy
artifacts of a certain type and the artifacts generated in their migration.
These artifacts are the following:~(i)~Forms Triggers and methods in the
generated managed bean class;~(ii)~Program Units and methods in the service
classes that where generated to implement program units, and
finally~(iii)~SQL code statements, that are present in both source and
target code. The second column of the table indicates the size of the
tested forms from which the metrics were obtained: small, medium, and
large. The rest of columns indicate the value of the metrics of the first
column for a particular artifact involved in the transformation chain:
(column~3) Oracle Forms; (column~4) KDM and Primitives models; (column~5)
Platform and Objectual models; and (column~6) Java application.

\begin{table}[!ht]
  \caption{Results\label{resultados}}
\begin{sf}
\begin{scriptsize}
\centering
\begin{tabular}{p{1.5cm} p{0.9cm} rrrr}
  \multicolumn{1}{>{\columncolor[gray]{0.90}}l}{{\textbf{\#elements}}} &
 \multicolumn{1}{>{\columncolor[gray]{0.75}}l}{{\textbf{ }}} &
 \multicolumn{1}{>{\columncolor[gray]{0.90}}l}{{\textbf{Forms}}}&
 \multicolumn{1}{>{\columncolor[gray]{0.75}}l}{{\textbf{KDM/Primitives}}} &
 \multicolumn{1}{>{\columncolor[gray]{0.90}}l}{{\textbf{Platform/OO}}} &
 \multicolumn{1}{>{\columncolor[gray]{0.75}}l}{{\textbf{Java}}} \\
  \hline
  \hline
  {\textbf{Triggers}} &   &          &     &        &               \\
  {\textbf{/ Methods}} &   &          &     &        &               \\
        & small         & 5        & 5        & 4           & 4    \\
        & medium         & 42        & 42     & 30       & 30   \\
        & large         & 251       & 251        & 246 & 246  \\
        {\textbf{Prog. Units}} &  &          &     &        &              \\
        {\textbf{/ Methods}} &  &          &     &        &              \\
        & small         & 25        & 25     & 25         & 25   \\
        & medium         & 30         & 30            & 30  & 30   \\
        & large         & 68         & 68     & 68        & 68   \\
        {\textbf{SQL sent.}} &   &          &     &        &               \\
        & small         & 13       & 13     &13            & 13   \\
        & medium         & 59        & 59          & 70  & 70   \\
        & large         & 169      & 169         & 200 & 200
\end{tabular}
\end{scriptsize}
\end{sf}
\end{table}

We have used ClearSQL for counting the triggers, program units, and SQL
code statements. We have also used this tool to find which triggers are empty.
Then we have subtracted these empty triggers from the total number of
triggers. Note that the existence of empty triggers is common in Oracle
Forms applications (mostly derived from human errors). The Metrics tool has
been used for counting the number of methods in Java classes.

With regard to the methods generated from the triggers migration, it is
worth noting that there are less methods than triggers. We have detected
that some triggers were missing during the injection because they were not
associated to an UI widget but they correspond to data block triggers.
Concretely,~1,~12, and~5~triggers were not generated for the small, medium
and large forms, respectively. In the case of program units, no methods
were missed.

The number of SQL code statements in Java code (and Object-oriented models)
is bigger than in Forms code and KDM/Primitives models. This is due to the
strategy adopted for migrating some SQL code statements. When a query code
statement includes a \texttt{SELECT \ldots{} INTO} structure for assigning
two or more columns into two or more variables, our solution splits the
original code statement in two or more new code statements where only one
column is projected into one variable.

\subsection{Limitations of the validation}

In this section we will enumerate some limitations of the carried out
validation.

\begin{enumerate}

\item We have not created unit tests for all the generated methods because this 
required a great effort for large and complex methods. 
The code covered is of~46\% (222~LOC) for the
  small size form,~49\% (687~LOC) for the medium size form, and~23\%
  (1065~LOC) for the large size form.
\item The three validated forms migth not cover all the elements of the
  PL/SQL grammar or all the potential artifacts in the Oracle Forms
  architecture.
\item Potential mistakes when contrasting behavior of the generated code
  against the requirements of the legacy application.
\item Potential mistakes in counting elements when a visual inspection
  is used for the EMF tree editor.
\item Potential errors can be found when counting triggers in Oracle Forms by
  using ClearSQL7 because an empty trigger is counted by the tool but not
  considered in our solution.
\end{enumerate}

\section{Evaluation\label{sec:Evaluation}}

Metamodels and model transformations are the two essential elements in
developing a MDE solution. In this Section, we will analyze the main issues
related to them in the context of our migration work. Regarding the
metamodels used, we focus on the KDM metamodel. On the other hand, we will
discuss how model transformations have been written and tested. In this
discussion, we will address issues as the visualization of the output
models, and the feasibility of applying a test-first programming approach.
Additionally, we will present a comparison carried between Java, ATL, and
QVTo as m2m transformation languages. Finally, we will comment two issues
of great interest in migration scenarios: how the horseshoe model has been
implemented and how the model traceability has been managed.

\subsection{Using KDM}

When OMG launched KDM~1.0 in~2007, the company Open Canarias decided to
adopt this metamodel, as it provided the level of abstraction appropriate
to represent source code of legacy applications, as an alternative to use
AST models or other existing metamodels, such as EGL proposed by IBM. Since
then, the company has used KDM to represent GPL code in all its
model-driven modernization projects. KDM has provided two main benefits to
the company:~(i)~flexibility to model statements written in an ambiguous
way, and~(ii)~ability to reuse visualization and code analysis tooling for
different GPLs. It is worth noting that Open Canarias contributed to the
implementation of the ASTM metamodel in Modisco~\cite{modisco-open}. This
company has developed infrastructure to create KDM injectors for GPLs, as
explained in Section~\ref{sec:injection}. This infrastructure has been
applied to create injectors for COBOL, ABAP, and PL/SQL.

Below, we detail the knowledge gained when using KDM to develop the tool
presented in this paper.

\paragraph{KDM compliance} The cost of creating the PL/SQL injector
was~3~months/man. The injector is L1 compliant KDM for Data domain but not
for Micro-KDM. A family of stereotypes was defined to represent PL/SQL
statements instead of using Micro-KDM, as explained in
Section~\ref{sec:injection}. This has not negatively influenced the
migration implementation. Whenever an \texttt{ActionElement} is processed,
the \texttt{name} attribute must be used to check if it represents a code
statement of the expected type. KDM models could then be used as input of a
tool~\cite{AFP-JISBD} developed to implement the AFP
specification~\cite{AFP}. This tool was developed during the migration
project to calculate the function points of Oracle Forms applications, but
could be applied to other languages.

\paragraph{KDM limitations to manage Data from Code} According to the
KDM specification, the elements of a package can have references to
elements in packages of a lower level but not vice versa, as the models are
constructed from the code to the more abstract viewpoints. This lack of
navigability can make writing model transformation difficult. The injector
used created references from \texttt{Action} elements to \texttt{UI}
elements, but not to \texttt{Data} elements. This caused problems in
writing model transformations that required access to the database schema.
We solved this issue by directly moving the PL/SQL code statements to the Java
code. References to \texttt{UI} elements are really needed as observed in
the \emph{primitives2platform} transformation. They allowed us to access to
\texttt{UI} elements from \texttt{CallableUnit} in order to know to which
visual component an primitive is associated, as described in
Section~\ref{platform_mm}.

\paragraph{Benefits and limitations of KDM to model software concerns}
\emph{Infrastructure} and \emph{Program Elements} layers are very useful to
represent legacy code. L0 compliant KDM models are more convenient than AST
or CST models to perform a reverse engineering process, and L1 compliant
Micro-KDM models promote the interoperability. We have been able to
experiment with these benefits in the work here presented. Data managed by
legacy applications can also be adequately represented by using the
\texttt{Data} package. However, we have not used this package in our work,
as explained above.

From our experience, the rest of level~1 KDM packages are too generic to be
useful as-is. The \texttt{UI} package is very limited to represent the UI
information required in software modernization tasks. For example, this
package only includes the generic \texttt{UIField} and \texttt{UILayout}
classes to represent information on widgets and layouts, respectively. KDM
should therefore be extended in order to represent different kinds of
widgets and layouts. However, the KDM extension mechanism is very limited
in practice as described in Section~\ref{sec:KDM}, and when using
stereotypes the interoperability among tools (one of the main benefits of
KDM) is lost.

Therefore, our understanding is that defining (or reusing a existing
metamodel) tailored to the domain of interest is a better alternative to
KDM extensions as also noted in~\cite{oscar2016-ist}. In fact, Open
Canarias decided to use IFML and UsiXML metamodels to represent the legacy
UI, as indicated above. In short, KDM models are an appropriate
representation to start a model-driven reverse engineering process, but
other metamodels must be defined (or reused) to abstract the information
managed throughout a re-enginering process. In the project described here,
we have devised the Primitives, Platform, and Object-Oriented metamodels.
It is worth noting that we have not used the metamodel defined
in~\cite{wcre11} to represent the execution flow of primitives since this
flow is represented in KDM models.

The size of the KDM models for the large case study is nearly~60Mb, while
the sum of the sizes of the other three models is~5Mb. This evidences that
KDM models compliant L0 or L1 level represent code at a very low level of
abstraction.

\paragraph{Learning of KDM} Two of the members of our team learned KDM, in
particular the \texttt{Infrastructure} and \texttt{Program Elements} layers
and the \texttt{Data} and \texttt{UI} packages. They had a solid background
in MDE and were able to acquire required KDM knowledge in~50~hours. This
number of hours includes the time devoted to prepare a report about the
injection process developed by Open Canarias. Writing this report required
understanding on how each PL/SQL statement was represented by means of
stereotypes. A tutorial previously elaborated by our group facilitated the
learning of KDM. It is worth noting the lack of publicly available KDM
tutorials.

\subsection{Writing model-to-model transformations in Java}

Software migration is an MDE usage scenario which involves complex
model-to-model transformations. In Morpheus, Java (and the EMF API) was
considered a better option than using model-to-model transformation
languages such as ATL~\cite{atl}, ETL~\cite{etl}, or QVT
operational~\cite{qvt}. This choice was motivated by several
reasons:~(i)~immaturity and lack of stability in some of the required
tooling (for example in the case of QVT),~(ii)~the uncertainty of continued
future support in the case of ATL, which may lead to a high technical debt,
and~(iii)~the high complexity of the transformations involved would demand
to write a large amount of imperative code. In fact, we had already used
Java instead of ATL or RubyTL~\cite{rubytl} in some recent reverse
engineering projects~\cite{oscar2016-ist}. Moreover, Open Canarias had
performed an internal survey on m2m transformation languages, and the staff
concluded the convenience of using GPL or domain specific languages
embedded into GPL.

\begin{figure*}[h]
\centering\includegraphics[width=0.8\linewidth]{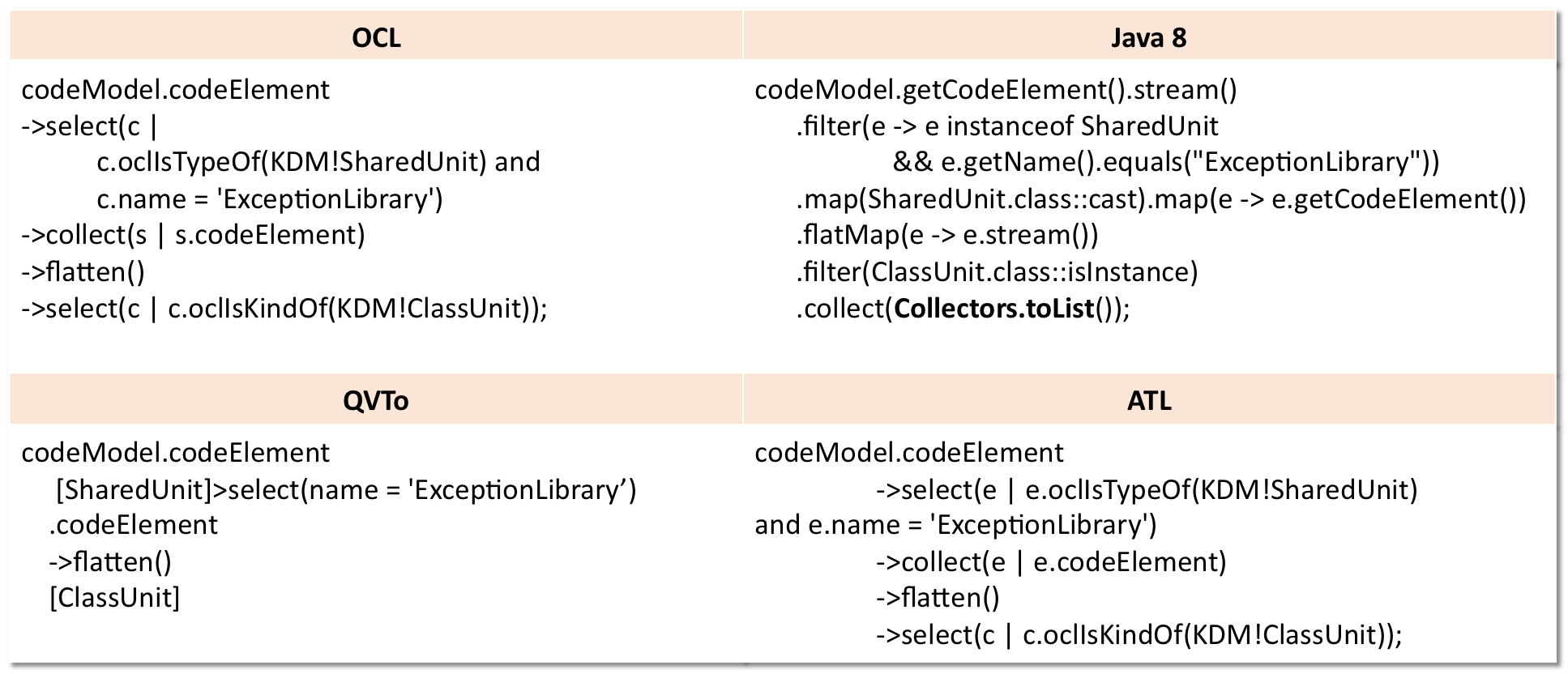}
\caption{Code comparison example of OCL, Java 8, QVTo and ATL.}
\label{fig:ocl}
\end{figure*}

Once the construction of the Code Migrator tool finished, we conducted a
study to compare Java with two common m2m transformation languages: ATL as
hybrid language, and QVT operational as imperative language. We used one of
the m2m transformations implemented as a case study. Since~2007, the
\emph{Model Transformation Contest} is held to compare m2m tools through a
case study raised in advance. This workshop has originated the publication
of some valuable comparative studies of m2m
languages~\cite{Jakumeit14,Rose14}. Our comparison contrasts with these
studies as follows:~(i)~the case studies are more complex and are taken
from a real project;~(ii)~KDM is the source metamodel; and~(iii)~the same
developer has written the transformations.

The three m2m transformations implemented have the following size measured
in lines of code (LOC):~1926 for \emph{kdm2dioms},~535 for
\emph{primitives2platform}, and~4568 for \emph{platform2oo}. We
therefore chose the \emph{kdm2primitives} transformation as case study because
had a medium size and complexity. Writing the ATL and QVTo transformations,
we considered the features of each language. ATL is a hybrid language that
allows to express the mapping between the source and target metamodels in a
declarative way, and provides imperative constructs to express more complex
parts of a transformation. QVT Operational is an imperative language that
is part of the QVT hybrid specification~\cite{qvt}, but that can be used
separately from the QVT relational language. Both ATL and QVTo provide
rules and helpers to express transformations. Helpers are used to factorize
code and achieve short and legible rules. The ATL rules have a special
clause to write imperative code that is executed after applying mappings.
Whereas ATL helpers can only include OCL expressions, QVTo helpers can
include any kind of statement and the state of the transformation can be
changed. In addition, QVTo allows to declare intermediate data by means of
\texttt{attribute}s. In ATL, the order in which rules are executed is
implicitly determined, whereas this order must be explicitly expressed in
QVTo code. Regarding the creation of target elements, this is implicit in
ATL but explicit in QVTo.

As explained in Section~\ref{sec:Development}, we have organized the
model-to-model transformations in four components: Iterator, Analyzer,
Builder, and Reference Resolver. How these components have been implemented
for the \emph{kdm2primitives} transformation was described in detail in
Section~\ref{sec:idioms}.

The ATL and QVTo transformations have been organized as follows. In both
cases, a rule has been defined for each mapping kdm-primitives.
Intermediate data are required to record symbol tables for local and global
variables and exceptions. These tables have been declared as properties
(\texttt{dictionary} type), whereas the dynamic map
pattern\footnote{Unofficial ATL Tutorial:
  \url{https://github.com/jesusc/atl-tutorial}.} %
has been applied in ATL. In ATL, we have encountered difficulties in
writing the filters needed to discriminate the kind of statement
represented by an \texttt{ActionElement} source element, and sometimes we
had to explicitly invoke rules. Using QVT properties to store intermediate
data instead of query helpers allowed us to reduce the execution time.

In ATL, the transformation has~37 declarative rules~(444~LOC),~5 rules
embedding imperative code~(85~LOC), and~14~helpers~(50~LOC). In QVTo, the
transformation has~41~rules (316~LOC),~11~helpers~(76~LOC),
and~8~properties. The Java transformation has 1926~LOC distributed among
the four components as follows: Iterator~(325~LOC), Analyzer~(148~LOC),
Builder~(1077~LOC), and Reference Resolver~(376~LOC).
It is worth noting that streams and lambda expressions in Java~8 provide an
expressiveness similar to the OCL language to navigate models. Therefore,
the effort devoted to write model navigation expressions could be
considered equivalent in the three languages. Figure~\ref{fig:ocl} shows
the code to count the number of exceptions in a KDM model for OCL, Java~8,
QVTo and ATL. However, the creation of target elements is tedious in Java
because factory classes and getter/setter methods must be used.

The size and development time for each transformation has been the
following:~579~LOC and~88~hours for ATL,~392~LOC and~48~hours for QVTo,
and~1926~LOC and~160~hours for Java. It should be noted that writing the
Java transformation required understanding the problem and designing the
solution, which approximately involved half of the time spent in the
implementation.

We have measured the performance of each transformation. Each
transformation has been executed five times for two different inputs: a
medium~(87~elements) and large~(190~elements) KDM model. The
transformations have been executed under a Core~i5 CPU at~2.7Ghz,~8Gb of
RAM and~3Mb of cache, a SSD hard disk, MacOS Sierra~10.13.3, ATL~3.8, QVTo
Eclipse~3.7 and Java~8. For each run, the execution time is calculated as
the sum of the completion times for three tasks: load the input model,
execute the transformation, and write the target model.

\begin{figure}[h]
\centering\includegraphics[width=0.95\columnwidth]{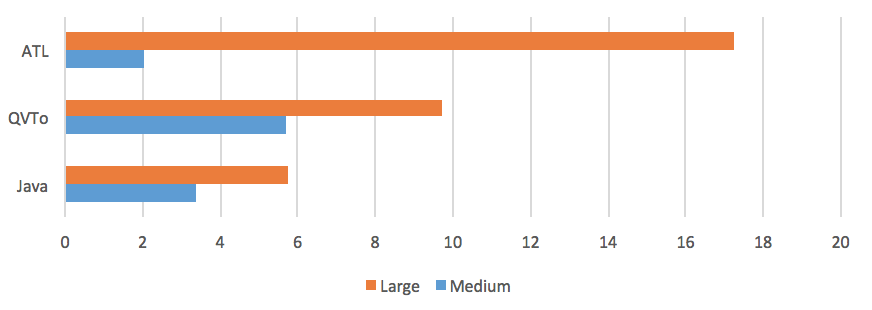}
\caption{Average execution times for the Java, ATL, and QVTo
  transformations.}
\label{fig:runningtimes}
\end{figure}

Figure~\ref{fig:runningtimes} shows the average execution times for each
input to the transformations. The variance obtained has been low in all the
cases with values in the range between $0.01$ and $0.06$.

The transformations have scaled out better for Java and QVT than for ATL.
The lowest execution time were for Java in the case of the large-size
model, and ATL for the medium-size model. It is worth noting that
performance of model transformations can significantly be reduced by
applying patterns for a particular language. For example, in the case of
ATL for the large-size model, we reduced the running time by half by using
the dynamic map pattern previously cited to store the symbol table for
variables and exceptions, instead of using \texttt{Dict} type attributes in
ATL, which are accessed from imperative code sections. ATL supports
parallel execution of transformations, and parallel streams of Java~8 can
be used to improve the performance of model transformations, but these
mechanism have not been explored.

Therefore, the results obtained for development effort and performance
would support the use of QVTo against either Java or ATL. However, the
existing tools for QVTo lack of the maturity offered by commonly used
tooling for software development. Software companies might prefer to
implement m2m transformation in Java to avoid using immature tools which
could be discontinued in the short term. While mature tools are not
available for m2m transformation languages, we think that the use of an
internal DSL (fluent-API or embedded DSL) could be the more appropriate
solution to write model-to-model transformations. In the case of Java, a
fluent-API to manage models could be useful. In fact, some tools have been
presented with this purpose~\cite{dsl2jdt,flapi}, although these projects
were discontinued. A tooling aimed to automate the creation of fluent-API
for the APIs that EMF generates could be very valuable to developers
writing m2m transformations in Java.

\subsection{Developing and testing model transformations}

Two essential aspects of the development process here applied are:~(i)~to
write transformations in a incremental way, and~(ii)~to apply different
kinds of tests for validating the transformation chain.

Writing complex transformations for large metamodels requires a systematic
approach. The transformations should be created and tested incrementally.
For this, we have defined a strategy similar to the test-first programming
as explained in Section~\ref{sec:Development} and illustrated in
Figure~\ref{fig:lifecycle}. The transformations are incrementally
developed, addressing one source-target mapping at a time.
In each step, an input model and the code that implements the mapping are
created.
That is, the models establish the order in which mappings are
implemented and tested. However, tests were not written to validate the
transformations. This decision was taken after writing some tests for the
\emph{kdm2primitives} transformation. We observe that test code would be very
similar to the transformation code to be tested.
Below we show the code of the test written to validate if the
\emph{kdm2primitives} transformation generates as many ``CASE''
\texttt{SelectionFlow}s that have as many \texttt{CASE} primitives as
\texttt{IF ELSIF ELSE} elements there are in the KDM model for the
corresponding PL/SQL code statements. JUnit was used to write and run the tests.
Although Java~8 facilitates the writing of the transformations and
significantly reduces the number of lines of code because the use of
streams and lambda expressions, writing tests still requires a considerable
effort.

\begin{lstlisting}[language=java]
for (SelectionFlow selectionFlow : selectionFlowList) {
  ActionElement actionElement = selectionFlow.getActionElement();
  // Get ELSE
  Optional<ActionElement> posibleElseActionElement =
      actionElement.getCodeElement().stream()
          .filter(c -> "ELSE".equals(c.getName()))
          .map(ActionElement.class::cast)
          .findFirst();
  if (posibleElseActionElement.isPresent()) {
    ActionElement elseActionElement = posibleElseActionElement.get();
    // Counts ELSIFs
    long elseIfs = elseActionElement.getCodeElement().stream()
          .filter(c -> "IF".equals(c.getName()))
          .filter(ActionElement.class::isInstance)
          .map(ActionElement.class::cast)
          .filter(a -> a.getStereotype()
                           .stream()
                           .filter(e ->"elsif".equals(e.getName().toLowerCase()))
                           .count() > 0)
          .count();
    // Check if ELSE has code
    long elsePresent = elseActionElement.getCodeElement().stream()
          .filter(c -> !"IF".equals(c.getName()))
          .count() > 0 ? 1 : 0;
    // IF + ELSE + ELSIFs
    assertEquals(1 + elsePresent + elseIfs, selectionFlow.getCase().size());
  } else {
    // IF
    assertEquals(1, selectionFlow.getCase().size());
  }
}
\end{lstlisting}

The creation of transformation-specific asserts could facilitate writing
tests. However, the implementation of these asserts is a challenging
problem since they should be applicable on any metamodel. We consider that
the solution proposed in~\cite{mcgill2007test} is not practical because XMI
format models must be navigated by means of XPath expressions~\cite{xpath}.
Therefore, writing unit tests before the transformation code, following the
test-first programming of agile methods, seems a practice not suitable in
the context of m2m transformations with the existing technology.

With regard to writing unit tests for m2t transformations, we consider
interesting the work of Tiso et al.~\cite{tiso14}. We could not use their
tool, as it is intended to be used on UML with profiles. However, the
approach can be adapted writing parameterized regular expressions specific
to our models, and the generation of Java code is also suitable for their
``sub-transformation'' approach following the containment relationships in
the target platform model.

We have combined unit tests for a particular model transformation with
tests aimed to validate the model transformation chain. As indicated in
Section~\ref{sec:Validation}, coverage and acceptance tests have been
applied along with syntactic and semantic tests on the transformation chain
output through out compiler tools.

In model-driven migration scenarios, test models can directly be obtained
from source code by using the injector. This is a clear advantage with
respect to other model-driven scenarios. Test models can be injected for
legacy code or either they can be injected from code specially written for
testing. In our case, we have used legacy forms provided by the company to
validate the model transformation chain, and have created small code
fragments to inject the input models used to validate the first m2m
transformation in the chain. The rest of m2m transformations were validated
by using the models generated in the precedence transformation according
the chain.

\subsection{Visualizing output models for testing model transformations}

Checking the correctness of the output model of a model transformation is a
complicated task due to the complexity of the metamodels involved. Either
creating oracles to be used with model comparison tools or performing
manual checking are really tedious and time-consuming tasks. Because the
difficulty of automatically or manually creating the expected output
models, a manual checking is usually performed. Input and output models are
examined to verify whether mappings between them are correct. This labor
can be facilitated by graphically visualizing output models, but the
generation of graphical editors tailored to each metamodel requires a
considerable effort. EMF provides a generic tree editor which shows a list
of all the model elements and allows navigating over the aggregation
hierarchy of each element, but references between objects are not
visualized. Examining models with this editor is really difficult.

\begin{figure}[!h]
  \centering
  \includegraphics[width=0.75\columnwidth]{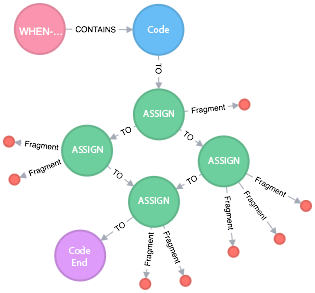}
  \caption{Neo4J graph for a model representing an execution flow.}
  \label{fig:neo4j}
\end{figure}

Because the models are graphs, we have studied how data visualization
capabilities offered by some graph databases can be used for visualizing
output models. We built a m2t transformation that generates Neo4J insertion
scripts from models representing the execution flow of primitives. These models
were finally not used because the execution flow is already represented in
the KDM models. The m2t transformation was easy to implement because the
execution flow models are graphs whose nodes are statements and the arcs
denote all the possible paths that might be traversed during the code
execution. The effort to write the transformation was only of about~4~hours.
Figure~\ref{fig:neo4j} shows the execution flow graph for the
PL/SQL code below:

\begin{lstlisting}[language=SQL,numbers=left,numberstyle=\tiny,morekeywords={BEGIN,IF}]
threshold := 1000;
IF salary > threshold THEN
  bonus := salary - threshold;
ELSE
  bonus := 100 + threshold - salary;
END IF;
salary := salary + bonus;
\end{lstlisting}

As observed in Figure~\ref{fig:neo4j}, the node color is used to
differentiate the executed trigger (\emph{pink}), the statement fragment
(\emph{green}), the initial node (\emph{blue}), and the final node
(\emph{purple}). This validation would have been very difficult by using
the EMF tree editor. However, it was easily performed through the graph
visualization. In addition, the Neo4J Desktop viewer allows that nodes included
in a graph can be shown on demand. In this way, the execution flow
can be easily followed, instead of facing us to a very large graph on the screen.

This approach of converting models in graph database objects is applicable
to many metamodels. It would require to establish a mapping between the
model information to be visualized and graph elements. In our case, the
only visualization carried out was for the execution flow models, which
were used in the final implementation. It might also have been used to show
references among the platform model elements. It allows us to check if
trigger code has been correctly spread in services (only one method service
or more). Another scenario could be to graphically represent the use of
variables for each method. In case two or more methods were referring the
same variable, then that variable will be tagged as shared. Therefore, it
would allow us to visually verify if the algorithm for sharing variables
worked. However, the fact of managing many small test models influenced in
no implementing more Neo4J-based visualizations.

It should be noted that Neo4J offers capabilities
that go beyond the visualization. Since it is a graph database,
Neo4J allows to perform queries that show a subgraph with elements satisfying
a particular condition. In our case, we have used this capability
to show the small graphs showing the control flow of a procedure.

\subsection{Model traceability}

Model traceability facilitates the implementation of unit tests in JUnit by
taking advantage of the existence of references from primitives to KDM model
elements. It allows us to integrate the developed tool into the one provided
by the company (UI2Java). These references enable backward navigation from
model elements resulting of transformations involved in the restructuring
stage. Therefore, both tools can be connected by means of the generated
models.

Model traceability also avoids having to decorate the generated models
after each transformation of the tool chain. When the Primitives model was
implemented, we decided to keep the back references in order not to add new
attributes on the elements for identifying the corresponding element type
in KDM. For instance, in a \texttt{SelectionFlow} element on the primitive
model, the reference to the corresponding KDM element provides the needed
information about if it was generated by an \texttt{If} or an
\texttt{Switch} KDM element. A specific attribute to identify the
corresponding provenance would have been included, but then the Primitives
metamodel would had been extremely overloaded until the fact that all the
KDM metamodel could be included inside it. It is worth noting that the
Primitives metamodel was devised to describe the data at a higher level of
abstraction than the KDM metamodel.
\section{Related Work\label{sec:RelatedWork}}

In this work we have presented a practical experience in developing a
model-driven reengineering approach for migrating legacy Oracle Forms code
to MVC platforms. We have defined a development process that integrated
several kind of validation, and present a strategy to implement model
transformations in a incremental way. Moreover, an assessment of applying
modeling techniques in this scenario is given. In this setting, the works
more closely related to our approach would belong to the following
categories:~(i)~model-driven approaches for migrating legacy
code,~(ii)~development processes for MDE solutions, and~(iii)~model-driven
migration experiences. In addition, we also consider studies on the use of
KDM, given this OMG metamodel is a key element in our solution.

Before of addressing related works in the above categories, it is worth
noting that the state of the art in model-driven reverse engineering has
recently been analyzed in~\cite{Raibulet2017}. The authors present a
systematic literature review in which they describe the more relevant
approaches and perform and analysis conducted by some research questions
related to the used metamodels and tools, and the level of automation
achieved. A total of fifteen proposals are detailed and some hints on
choosing a model-driven reverse engineering approach are pointed out.

\paragraph{Model-driven approaches for migrating legacy code}

A model-driven reverse engineering approach to migrate RAD event handlers
to modern platforms is proposed in~\cite{wcre11}. An AST model is injected
from legacy code, and a reverse engineering process is applied to obtain a
more abstract representation. Idioms models are firstly extracted, and then
an execution flow model is generated. This model represents a graph whose
nodes are idioms fragments and the arcs denotes the execution flow.
Additionally, each fragment is annotated to indicate which concerns it
belongs to: UI, Control or Business Logic. As proof of concept, Ajax code
was generated from execution flow graphs. Concretely, PL/SQL event handlers
were migrated to a two tier architecture: a HTML/Javascript user interface
invokes a REST service that implements business logic fragments. As noted
in Section~\ref{sec:Implementation}, we have used abstracted code into
Idiom models, but the reverse engineering here presented differs in two
significant aspects from the previous work:~(i)~KDM is used for
representing legacy code, and the execution flow model is not necessary
because that information is part of KDM models,~(ii)~separation of concerns
is achieved through a Plaform model, which is used to conduct the
object-oriented code generation, and~(iii)~new primitives have been identified
aims to model in KDM all the PL/SQL code statements.

In~\cite{andrade2008} a strategy for reengineering legacy systems to SOA
({\em Service Oriented Architecture}) is described. The authors consider
the decomposition of a monolithic application into several concerns as the
main problem to be addressed. They propose manually annotating legacy code
to indicate which tier each statement belongs to: logic, data, and user
interface. The reengineering process is defined according to the horseshoe
model. An AST model is injected from the legacy code, and then a
graph-based transformation language is used to redesign the code for the
new platform. The authors outline two strategies that they are considering
to generate code from the graph model obtained.

Unlike the works of Sanchez et al.~\cite{wcre11} and Heckel et
al.~\cite{andrade2008}, we have tackled all the stages of a reengineering
process. Beyond presenting a reengineering approach, we have defined and
applied a development process, which allowed us to explore some key aspects
of MDE solutions, such as writing and testing model transformations, and
the visualization of models with the purpose of testing. Our reengineering
solution has been applied to legacy code from real applications, and have
carried out several kinds of testing for validating the result.

\paragraph{Development processes for MDE solutions}

In~\cite{kuster2009}, the authors propose an approach for providing
incremental development of model transformation chains based on automated
testing. The approach includes four test design techniques along with a
framework architecture in order to test transformation chains. The paper
also includes a validation of the approach by developing a transformation
chain for model version management (for the IBM WebSphere Business
Modeler). As it is indicated, doing testing in an isolated way is not
sufficient, and the software quality should be assured by means of a
development process. They introduce the requirements to ensure the quality
of a transformation chain:~(i)~an iterative and incremental
development,~(ii)~testing processes for the transformation chain,
and~(iii)~a fully automated test environment. The incremental development
is based on automated testing, so its approach is supported by a test
framework which follows the TDD (test-driven development) principle. The
three testing design techniques that the proposal distinguishes are:
integrity tests on generated model, comparing the result of a model
transformation with inspected reference outcomes, invariant validation, and
deviation testing that consists on calculating and storing some data on
output models with the purpose of be comparing them with the data obtained
when the transformation is again executed.

We have applied an incremental development process similar to that proposed
by Küster et al.~\cite{kuster2009} to develop a reengineering based on the
horseshoe model. However, while that process is focused on transformation
chains, we have considered how unit tests can facilitate the implementation
of individual model transformations. Small input models, which cover one or
a few statements of the source language, are used as test models. We have
shown how this technique has been applied in our reengineering. Instead, an
incremental approach is proposed in~\cite{kuster2009} but its application
is not illustrated with a case study. Conversely, we have performed
acceptance tests and unit testing on the generated code. Therefore, we have
combined manual and automated testing. An automated test framework for
model transformations is desirable but its development is beyond the scope
of our work. As far we know, there are no automated testing frameworks for
model transformations publicly available today.

A notion similar to the unit tests proposed in our approach was proposed by
D. Varró in~\cite{Varro}, where the \emph{model transformation by example}
technique is presented. This technique aims to semi-automatically generate
model-to-model transformation rules from a collection of interrelated input
and output models. Each pair of this collection acts as prototypical
instance that describes a critical case to be addressed by the
transformation. Therefore, our unit test strategy puts into practice a MBTE
process. Varró et al. consider that ``the majority of model transformations
has a very simple structure,'' and transformations could therefore be
partially synthesized. However, we managed a very large metamodel as KDM,
and complex transformations, and it is not clear that the statement holds
for our case, so a manual writing and testing has been carried out.

As commented in Section~\ref{sec:Evaluation}, the work by Tiso et
al.~\cite{tiso14} is applicable to our work as a way of testing
model-to-test transformations. In their work, they divide the
transformation in sub-transformations, following the containment
relationships of the source metamodel. For each transformation, they build
parameterized regular expressions that capture the results--in form of text
outputs--of this transformation, but ignore the results of the
sub-transformations. This allows to effectively associate a regular
expression to the production of each of the transformations, allowing a
systematic test of the output text. The approach can be used in our
development, but their tool could not be used, as it is based on UML and
profiles, while we use EMF/Ecore, but a similar approach could be
considered in future refinements of the process.

A test-driven development (TDD) approach for model transformation is
considered in~\cite{jemtte}. Concretely, the JUnit framework is extended in
order to facilitate the application of TDD in developing model
transformations. Some examples of unit tests are shown for the built
prototype. From our experience, writing these unit tests requires a
significant effort because they have a large amount of code, and even the
correctness of this code should be validated. Therefore, we have not
written such unit tests. An interesting future work would be the
development of a unit test framework for implementing model transformation
in Java. EUnit~\cite{eunit} is an integrated unit test framework for model
management tasks that is based on the Epsilon platform. Unit tests are
written in the EOL language provided by Epsilon. Writing EOL unit tests was
not considered because our transformations were written in Java.

A model-driven software migration methodology is proposed
in~\cite{wagner2014model}. This proposal has been validated by means of a
case study based on a refactoring of a control software for wind tunnels
programmed in C/C++. In the reverse engineering stage a parser is used to
generate an annotated parse tree which is translated into XML format. This
format is imported into the jABC modeling tool to create customized process
graphs to obtain a code model (the control flow graphs of the application).
To validate the code models they use a back-to-back test. Firstly, the
original source code is restored from the code models by using a code
generator, then it is compiled. Thereafter, the restored code undergoes a
second reengineering step that includes to restore the source code again
from the new code model, and comparing both source codes. In the forward
engineering stage, the UI is analyzed to identify all possible actions of
the application. An abstraction model is created with the graphical
elements linked to their implementation. The migration and remodeling of
the application is a manual task. With the help of the abstraction model,
the developer examines the source code to identify and migrate the
necessary objects to a new process model. Finally, target code is generated
from the process model, but some code is also manually written. The
application is validated against the original one by comparing their
outputs. The two main differences with our approach are:~(i)~KDM is not
used to represent the legacy code, but control flow graphs obtained by
importing the XML code generated by the parser,~(ii)~the reverse
engineering stage is a manual task: The developer examine the code model
(control flow graphs) to identify and migrate the necessary objects. At the
forward engineering stage some code must be manually written as
well,~(iii)~the process of construction and testing of model
transformations is not addressed, and~(iv)~only a kind of validation is
applied: code generated is injected and compared with the refactored source
code.

\paragraph{Model-driven reeengineering experiences}

In~\cite{fleurey07}, one of the first model-driven reengineering
experiences is reported. The work presents the migration of a large scale
bank system to the J2EE platform. The three stages of the process are
commented and some conclusions are drawn. The model-driven approach is
compared to a manual process and some benefits and limitations are
provided. However, no details are provided on the development process
followed, and the transformation chain is not explained in detail. The
models are injected in form of an AST tree.

A model-driven approach for applying a white-box modernization approach has
recently been presented in~\cite{garces2017white}. First, a technology
agnostic model is obtained from the sources. Then, this model is edited by
the developer to configure the target architecture, and finally a
transformation from the model into the new technology is performed. They
used an Oracle Forms migration to Java technology case study, and were able
to generate the graphical interface (but without layout), the logic related
to database operations (e.g. read, update), and the scaffolding code to
call the PL/SQL logic that are embedded in the triggers. In contrast to our
work, the PL/SQL logic is not transformed and must be manually migrated.

In~\cite{Ebert2011}, most widely used reengineering tools are contrasted.
In this study some model-driven frameworks are considered as Blue Age~\cite{blueage},
Modisco~\cite{modisco}, and Moose~\cite{moose}.

\paragraph{Use of KDM}

As can be noted in~\cite{Raibulet2017}, a few experiences on the use of KDM
have been reported. Below, we comment two of those described in that
survey.

Pérez Castillo et al.~\cite{marble} propose a method for recovering
business processes from legacy information systems using MARBLE
(Modernization Approach for Recovering Business processes from LEgacy
Systems).
This method considers~(i)~a static analysis to extract knowledge from the
legacy source code,~(ii)~a model transformation based on QVT to obtain a
KDM model from that knowledge, and~(iii)~another QVT transformation to
create a business process model from the KDM model. MARBLE keeps
traceability because it identifies which pieces of legacy code were used to
obtain the elements of the business process.

Normantas and Vasilecas~\cite{Normantas2012} present an approach to
facilitate business logic extraction from the knowledge about a existing
software system. They use various KDM models at different abstraction
levels to represent the information extracted from the system in three
steps:~(i)~a preliminary study, to gather information about the
system,~(ii)~knowledge extraction into various KDM representation models,
and~(iii)~separation of the KDM model parts of the business logic from the
infrastructure ones. The authors apply source code analysis techniques to
identify the business logic and represent it with the KDM Conceptual model.

How ASTM and KDM models can be used to automate a modernization task was
addressed in~\cite{canovas2010}. Some lessons learned on the application of
KDM for calculating metrics for PL/SQL code were exposed.

Here, we have detailed discussed complex model transformations involving
KDM models, and have discussed some benefits and limitations of KDM from
our experience. Moreover, we have experienced the use of stereotypes
instead of injecting Micro-KDM compliant models. Both~\cite{marble}
and~\cite{Normantas2012} are not focused on the \texttt{Action} and
\texttt{Code} packages, but business process models are considered.
Finally, it is worth noting that we have here addressed a real
source-source migration process, while a simpler modernization problem was
considered in~\cite{canovas2010}.

\section{Conclusions and Future Work\label{sec:Conclusions}}

Model-driven techniques emerged early this century as the new software
engineering paradigm to achieve levels of productivity and quality similar
to other engineering areas.
Much attention has been devoted to the application of MDE in engineering
scenarios, both in the industry and in the academy. As recently noted
in~\cite{Raibulet2017}, the literature on the application of MDE techniques
in model-driven reverse engineering and re-engineering is very limited.
Practical experiences such as the one described here can be useful to know
the benefits and limitations of model-driven techniques in such scenarios.
Moreover, they can contribute defining new processes, techniques, and
practices, or just experimenting with the existing ones.

In this paper, we have presented a practical experience in designing and
implementing a re-engineering approach for a migration of Oracle Forms code
to a MVC architecture based on Java frameworks. For this aim, we have
defined a systematic process aimed to the development of the model
transformation chain implementing the reengineering. This process involved
several kinds of tests for validating the model transformations and the
generated code. The definition of this process and the description of its
applications is a contribution with respect to previously published
model-driven reengineering approaches. Specially, we would remark the
strategy applied to incrementally develop model transformations.

Regarding the benefits of applying MDE in a software modernization
scenario, it is well known that~(i)~metamodels provide a formalism more
appropriate than other metadata formats (e.g. XML, JSON) to represent
information harvested in applying reverse engineering,
and~(ii)~transformations allow migration tasks to be
automated~\cite{ase10,Raibulet2017}. In this work, we also contribute with
knowledge about some specific concerns in a software migration. The main
contributions would be the following.

\paragraph{Use of KDM} Experimenting with KDM in a real project, and
showing that L1 compliant KDM models for data and Micro-KDM are very
convenient to represent code from a structural and behavioral viewpoint.
Thanks to the usage of KDM, we did not need to define a PL/SQL, Data or
execution flow metamodels. In addition, KDM provides a code and data
representation devised by experts in migration. This a clear example on the
advantages of asset reusing.

\paragraph{Writing model transformations} We have compared Java with two of
the most widely used languages for writing model-to-model transformations.
This comparison was performed for one of the transformations of our work.
We have concluded that a fluent-API for Java (backed by EMF/Eclipse as the
more widespread modeling framework) is the more appropriate choice until
mature and robust environments for model-to-model transformation languages
are available.

\paragraph{Models for testing} Creating input and expected output models
for validating model transformations is very difficult as noted
in~\cite{baudry2010}. We have shown that software migration makes it easier
to have input models as they can be achieved from source code by using the
model injector. On the other hand, the model transformation chain generates
code. Checking the correctness of this code, and its correspondence with
the original code to be modernized, is again a very difficult task. The
usage of a test-first, incremental transformation development helped in
checking each of the mappings from models to text. Other approaches were
also considered and left as future work~\cite{tiso14}. It is worth noting
that we have investigated the transformation of models into instances of
graph databases in order to take advantage of viewers and query languages
offered by these database systems. This would provide a simple strategy to
visualize output models of a transformation, which would facilitate the
manual validation.

The practical experience here presented has served to define some research
works for some issues here raised, such as:~(i)~extend JUnit to develop a
automated unit test framework for model transformations,~(ii)~explore
test-driven development for model transformations,~(iii)~complete the
comparison among transformation languages,~(iv)~integrate
\emph{Models4Migration}~\cite{ruiz17} with the model management
platform of Open Canarias, and~(v)~to define a systematic approach to
visualize models as graph databases and explore the benefits of this
representation for automating different kinds of testing (e.g. coverage
testing).

\bibliographystyle{elsarticle-num}
\bibliography{main}

\end{document}